\documentclass[aps,prd,reprint,groupedaddress, onecolumn,nofootinbib]{revtex4-2}

\usepackage[utf8]{inputenc}
\usepackage[english]{babel}
\usepackage[T1]{fontenc}
\usepackage{amsmath,amsfonts,amsthm,bm}
\usepackage{amsfonts}
\usepackage{hyperref}
\usepackage{amssymb}
\usepackage{graphicx}
\usepackage[left=2cm,top=2cm,bottom=2cm,right=2cm]{geometry}
\usepackage{subfig}
\usepackage{physics}
\usepackage{array}
\usepackage{tcolorbox}
\usepackage{aas_macros}
\usepackage{multirow}

\newcommand{\msun}{\text{M}_\odot}

\newcommand{\hie}{hierarchical }

\begin{document}



\title{Transverse Doppler effect and parameter estimation of LISA three-body systems}


\author{Adrien Kuntz}
\email[]{adrien.kuntz@sns.it}
\affiliation{Scuola Normale Superiore, Piazza dei Cavalieri 7, 56126, Pisa, Italy}
\affiliation{INFN Sezione di Pisa, Largo Pontecorvo 3, 56127 Pisa, Italy}
\author{Konstantin Leyde}
\email{kleyde@apc.in2p3.fr}
\affiliation{Universit\'e Paris Cit\'e, CNRS, Astroparticule et Cosmologie, F-75006 Paris, France}


\date{\today}

\begin{abstract}
Some binary black hole systems potentially observable in LISA could be in orbit around a supermassive black hole (SMBH). The imprint of relativistic three-body effects on the waveform of the binary can be used to estimate all the parameters of the triple system, in particular the mass of the SMBH. We determine the phase shift in the waveform due to the Doppler effect of the SMBH up to second order in velocity, which breaks a well-known exact degeneracy of the lowest-order Doppler effect between the mass of the SMBH and its inclination. We perform several parameter estimations for LISA signals including this additional dephasing in the wave, showing that one can determine accurately \textit{all} the parameters of the three-body system. Our results indicate that one can measure the mass of a $10^8\,$M$_{\odot}$ SMBH with an accuracy better than $\sim 30\%$ (resp. $\sim 15\%$) by monitoring the waveform of a binary system whose period around the SMBH is less 100 yr (resp. 20 yr).

\end{abstract}


\maketitle

\section{Introduction}

Detection of gravitational waves (GWs) from compact binaries by the LIGO-Virgo-KAGRA collaboration has now become a routine process, with as much as 35 signals in the second part of the third observing run~\cite{LIGOScientific:2021djp}. Scenarios for the formation of these compact binary black holes (BBHs) include evolution of isolated stellar binaries~\cite{Belczynski_2004, Dominik_2012, Kinugawa_2014}, dynamical formation in dense stellar clusters~\cite{Portegies_Zwart_2000, Rodriguez_2016, Antonini_2016, Antonini_2016_2} or in active galactic nuclei (AGN)~\cite{Bellovary:2015ifg, Bartos_2017, Stone_2016, Tagawa_2020}.
Some of these events could also be observed in the future space-based interferometer LISA years before they merge in the LIGO-Virgo band~\cite{Sesana_2016, Toubiana_2020}. Observation of the signal during the entire 6 years of the recommended LISA mission duration will allow for an exquisite measurement of the parameters of the system~\cite{Toubiana_2020}. In particular, any environmental effect could be detectable if it induces a large enough phase shift~\cite{Caputo_2020, Barausse_2014, Barausse_2015,Cardoso_2020}.

A typical example of these environmental effects is the presence of a distant third body, called perturber, in the vicinity of the BBH (such systems are called "hierarchical", in the sense that the motion can be split between the inner orbit of the BBH and the outer orbit of the perturber). Indeed, three-body systems are quite common in the Universe: for example, 90\% of low mass binaries with periods shorter than 3 days are expected to belong to some hierarchical structure~\cite{Tokovinin_2006}, and we have now observed a pulsar in a triple system~\cite{2014Natur.505..520R}. Measurement of the parameters of the perturber from the waveform could provide relevant insights into the formation and evolution process of these BBHs, for example by allowing to determine if the BBH lies in a nuclear star cluster~\cite{Martinez_2020, Zevin_2019} or in the vicinity of an AGN~\cite{Toubiana_2021, 2022arXiv220508550S, Bartos_2017}. This last possibility is in fact particularly relevant for LISA detections, as the presence of "migrations traps" in accretion disks~\cite{Bellovary:2015ifg} around supermassive black holes (SMBH) implies the existence of a population of BBHs detectable by LISA~\cite{2022arXiv220508550S}, with 1 to 10 observable events during the mission duration. 
Additionally, the work of \cite{Yang:2019okq} predicts that between 4$\%$ and 40$\%$ of LIGO-Virgo detections could originate from binaries in AGN environments. 
In fact, it has been proposed that the relatively massive BBH GW190521 observed by LIGO-Virgo has formed in an accretion disk around an AGN~\cite{LIGOScientific:2020ufj}, consistent with the fact that the Zwicky Transient Facility reported an electromagnetic counterpart to this event~\cite{PhysRevLett.124.251102, 2021arXiv211212481C}. If indeed LISA detects binaries in accretion disks close to AGN, several environmental effects could be measured in the waveform: disk-induced migration and mass accretion~\cite{Kocsis_2011, Caputo_2020,Barausse_2014, Barausse_2015}, relativistic three-body resonances~\cite{Kuntz:2021hhm} or Doppler effect due to the motion of the center-of-mass of the BBH around the SMBH~\cite{Tamanini_2020, Inayoshi:2017hgw}. Measurement of the parameters of the SMBH from the waveform could be of valuable importance since other astrophysical probes suffer from several uncertainties and are often limited to sub-populations of SMBH~\cite{2014SSRv..183..253P}.

The Doppler effect is usually thought to be the largest, and numerous works studied in detail how the longitudinal Doppler shift induced in the GW phase of the BBH can allow to estimate the parameters of the third body~\cite{Robson_2018, 2019MNRAS.488.5665W, Randall_2019, Bonvin_2017, Inayoshi:2017hgw, https://doi.org/10.48550/arxiv.2109.08154, Toubiana_2021, 2022arXiv220508550S, 2019PhRvD..99b4025C}. Unfortunately, the longitudinal Doppler effect suffers from an intrinsic degeneracy between mass of the perturber and inclination (the same degeneracy being present in exoplanet mass determination by radial velocities~\cite{Wright_2018}). This can be intuitively understood from the fact that this effect involves the projection of the velocity of the center-of-mass of the BBH along the line-of-sight: a very massive third object viewed in a quasi-perpendicular configuration gives the same contribution as a less massive object parallel to the line-of-sight (see Fig.~\ref{fig:Osculating} for an illustration). In order to break this degeneracy, it was proposed to include higher-order effects in the waveform stemming from the relativistic influence of the SMBH: de Sitter precession of the angular momentum of the BBH due to spin-orbit coupling~\cite{Yu_2021}, Kozai-Lidov oscillations~\cite{Chandramouli:2021kts}, or Shapiro time delay during the propagation of the GW~\cite{2022arXiv220508550S}.

In this article, we will explore yet another way of breaking the degeneracies of the longitudinal Doppler shift: the transverse Doppler effect, which depends on the absolute value of the center-of-mass velocity and not only on its projection along the line-of-sight. Technically speaking, this effect is higher-order than the longitudinal Doppler shift (it is quadratic rather than linear in the center-of-mass velocity), however it can be shown to be of greater magnitude than both the spin-orbit precession or Shapiro time delay effects discussed above (see Fig.~\ref{fig:ParamSpace}). We also take into account in our analysis the gravitational redshift due to the potential well of the outer object (which we loosely include in our definition of "transverse Doppler effect"), which has the same order-of-magnitude in terms of post-Newtonian power-counting.
We will show that including this transverse Doppler effect in the waveform breaks degeneracies, allowing to measure \textit{all} parameters of the outer orbit of the perturber. In contrast, studies using only the longitudinal Doppler shift were able to measure only a subset of these parameters. We will carry out parameter estimation for typical BBHs in the vicinity of AGN using a Monte-Carlo Markov Chain (MCMC) algorithm \cite{Metropolis:1953am, Hastings:1970aa}, which will allow us to precisely study uncertainties and degeneracies among parameters.
Moreover, we will show that our method can allow to determine the mass of a $10^8\msun$ perturber with a precision better than 30\% up to periods of 100 years, corresponding to a distance of the BBH to the AGN of 0.05 parsecs. For smaller periods, we can reach a determination of the mass up to $14\%$ accuracy. 

This paper is organized as follows: in Section~\ref{sec:DopplerPhase} we will introduce our parametrization of three-body systems and derive the expression of longitudinal and transverse Doppler shifts in the GW phase, both in the time-domain and in the frequency-domain using a stationary phase approximation. In Section~\ref{sec:qualitative} we will give a qualitative analysis of the formula giving the transverse Doppler effect. In particular, we will explore the remaining degeneracies in limiting cases, and we will derive an observability criterion. In Section~\ref{sec:MCMC} we will explain our methodology for parameter estimation via Monte-Carlo Markov Chains, and in particular we will study the maximal distance of the BBH to the AGN up to which it is possible to measure the AGN mass with a reasonable accuracy. We will conclude with a discussion in Section~\ref{sec:conclusions}.
Throughout this article, we will work in units where $c=1$ and use the $(-+++)$ sign convention for the metric, while Newton's constant is denoted by the symbol $G_N$.

\section{Doppler phase factor in frequency-space} \label{sec:DopplerPhase}

\subsection{Three-body systems and osculating elements} \label{sec:osculating}

\begin{figure}
\includegraphics[width=0.25\columnwidth]{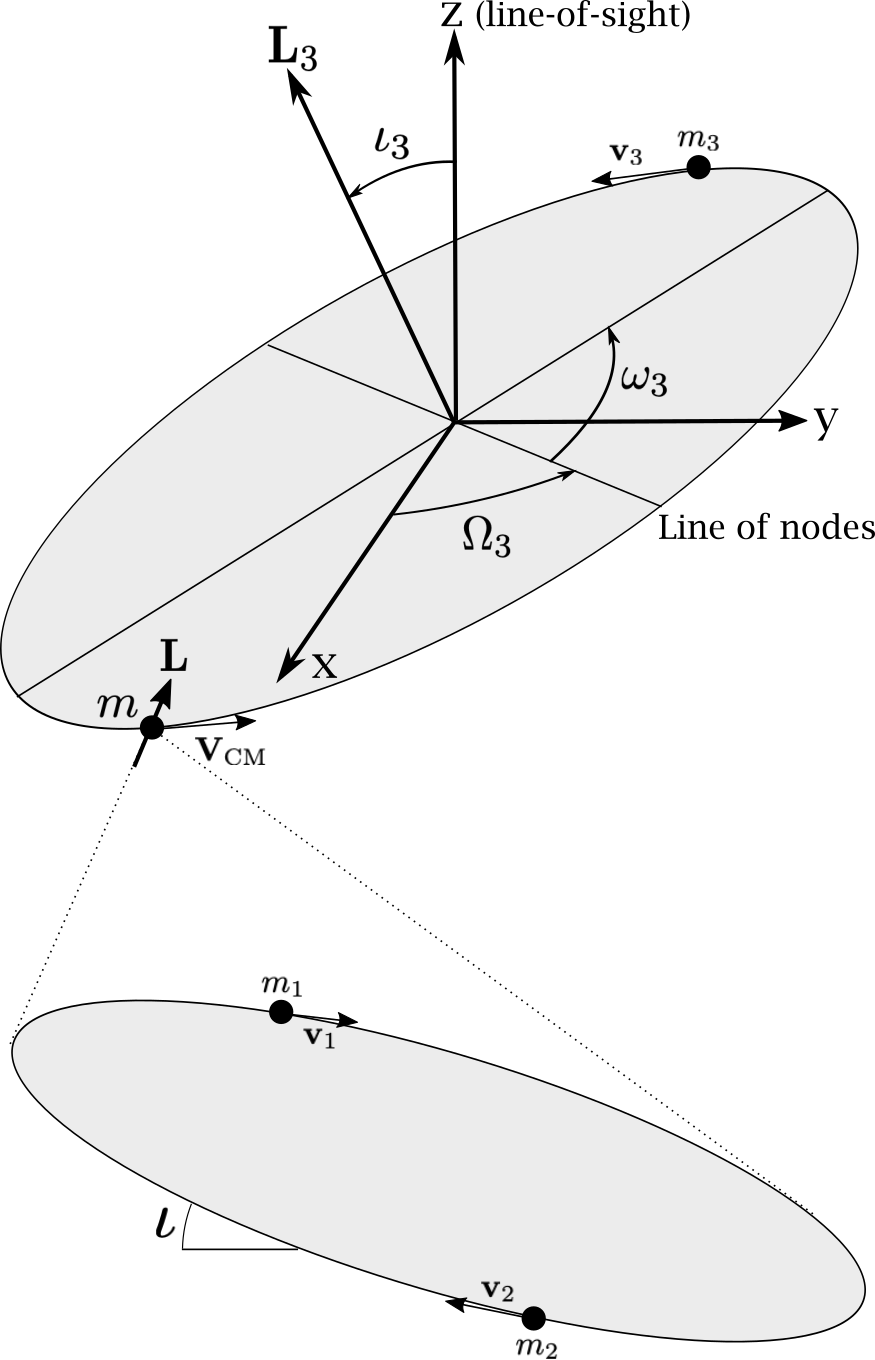}
\caption{Geometry of both inner and outer orbits, where $\bm L$ and $\bm L_3$ are the angular momentum vectors of the inner and outer binaries respectively.
}
\label{fig:Osculating}
\end{figure}

In this section we will set up our conventions for describing three-body systems and briefly discuss which parameters we can in principle extract from the waveform. We consider a \hie system of three bodies constituted by a tightly bound inner binary BH of masses $m_1, m_2$ whose center-of-mass orbits a distant perturber of mass $m_3$, and we denote by $m=m_1+m_2$ the mass of the inner binary and $M=m_1+m_2+m_3$ the total mass of the system. For the sake of generality, we will not assume that $m_3 \gg m_1, m_2$ when deriving the expression of the transverse Doppler terms, so that our computations are valid as well if the perturber is not a SMBH.
In a frame centered on the total center-of-mass of the three-body system, the positions of the three BHs are denoted by $\bm y_1$, $\bm y_2$ and $\bm y_3$, and we introduce the center-of-mass of the inner binary as $m \bm Y_\mathrm{CM} = m_1 \bm y_1+m_2 \bm y_2$.
 For simplicity, we assume all BHs to be nonspinning. We can decompose the motion into two ellipses osculating the trajectories, called inner (resp. outer) orbit, of period $P$ (resp. $P_3$). Each ellipse is characterized by a set of orbital osculating elements: for the inner orbit, these are the semimajor axis $a$, eccentricity $e$, initial phase $\varphi$, argument of perihelion $\omega$, inclination $\iota$ and longitude of ascending node $\Omega$ (resp. $a_3,e_3,\varphi_3,\omega_3,\iota_3,\Omega_3$ for the outer orbit), see Fig~\ref{fig:Osculating}. Of course, one has the equations $P = 2 \pi \sqrt{a^3/(G_N m)}$ and $P_3 = 2 \pi \sqrt{a_3^3/(G_N M)}$ relating periods to masses and semimajor axes. The radius vectors of the two orbits are denoted by $\bm r = \bm y_1 - \bm y_2$ and $\bm R = \bm Y_\mathrm{CM} - \bm y_3$ respectively, and are given as a function of planetary elements as
\begin{equation}
\mathbf{r} = a \left( (\cos \eta - e) \; \bm{\alpha} + \sqrt{1 - e^2} \sin \eta \;  \bm{\beta} \right) \; , \quad  \bm{\alpha} = R_z(\Omega) R_x(\iota) R_z(\omega)  \mathbf{u}_x \; ,  \quad \bm{\beta} = R_z(\Omega) R_x(\iota) R_z(\omega)  \mathbf{u}_y \; ,
\end{equation}
where $\mathbf{u}_i$ and $R_i$ are a unit vector and a rotation matrix along the $i$ axis respectively, and $\eta$ is the eccentric anomaly defined by Kepler's equation $\eta - e \sin \eta = 2 \pi (t-t_c) / P + \varphi$. Here we have chosen to use the time at coalescence $t_c$ as reference time, so that the angle $\varphi$ is the binary phase at coalescence.
Analogous formulas hold true for the outer orbit vector $\bm R$.
Finally, the three position vectors of the BHs $\bm y_1$, $\bm y_2$ and $\bm y_3$ can be expressed with $\bm r$ and $\bm R$ as
\begin{equation}
\label{eq:CM}
\bm y_1 = X_3 \bm R + X_2 \bm r \; , \quad \bm y_2 = X_3 \bm R - X_1 \bm r \; , \quad \bm y_3 = - X_\mathrm{CM} \bm R
\end{equation}
where $X_1 = m_1/m$, $X_2 = m_2/m$, $X_3 = m_3/M$ and $X_\mathrm{CM} = m/M$ are the mass ratios of both orbits.

In the literature on three-body systems, it is customary to assume that the $z$ axis of the coordinate system is aligned with the total angular momentum (the so-called "invariable plane"~\cite{valtonen_karttunen_2006,10.1093/mnras/stt302,Naoz_2016}). However, here we find it more convenient to orient the $z$ axis along the direction of the observer. 
We will assume that the inner orbit is circular and emits GWs in the frequency band of LISA. Then, it turns out that the only inner osculating parameters entering into the waveform of the inner binary are $a$, $\varphi$ and $\iota$~\cite{Maggiore:1900zz}. On the other hand, the additional Doppler shift induced by the motion of the center-of-mass of the inner binary \textit{a priori} depends on all the parameters of the outer orbit and of $m_3$, which adds seven additional parameters to the waveform. As we will see in Section~\ref{sec:DopplerDerivation}, both longitudinal and transverse Doppler shifts do not depend on $\Omega_3$, which reduces the additional parameters to six. So, in total, the waveform will be described by 15 parameters: 9 to describe the intrinsic waveform of the inner binary ($m_1$, $m_2$, $\varphi$, $\iota$, its distance $d_L$, the time to coalescence $t_c$, the angles $\beta, \lambda$ to describe the location of the system in the sky and the orientation of the plane of polarization $\psi$), and 6 parameters in the Doppler shift, which we take to be $a_3$, $P_3$, $e_3$, $\iota_3$, $\omega_3$ and $\varphi_3$~\footnote{Instead of using $P_3$ and $a_3$ as free parameters, one can use $P_3$ and $m_3$ as well; the former are more convenient for discussing the degeneracies among parameters, while the latter are more interesting for astrophysics purposes. We will switch from one set of parameters to the other depending on their convenience.}. We will study in Section~\ref{sec:qualitative} the degeneracies present in the additional parameters describing the Doppler shift.

An important remark here is that we will assume that all osculating parameters of both orbits are constant throughout the observation time, apart the inner semimajor axis $a$ which decays due to radiation-reaction. Of course, in generic three-body systems the osculating elements evolve over time (the Kozai-Lidov oscillations are a well-known example~\cite{LIDOV1962719, 1962AJ.....67..591K}). What we assume here is that this evolution occurs on a timescale long enough so that its contribution to the waveform is negligible. In some situations, this assumption could be violated, see e.g.~\cite{Chandramouli:2021kts,Yu_2021} for interesting examples on how to include osculating elements variations into the waveform. Note however that these effects are higher-order with respect to the transverse Doppler effect to which this article is dedicated: in the Lagrangian formulation of~\cite{Kuntz:2021ohi}, the three-body terms inducing variation of osculating elements (as e.g. precession of the inner binary angular momentum) are suppressed by a factor of at least $(a/a_3)^{3/2}$ with respect to the transverse Doppler terms (dubbed 'monopole' there). Thus, they should give a relevant contribution to the waveform only in a smaller section of parameter space, as we will illustrate in Figure~\ref{fig:ParamSpace}.

\subsection{Doppler effect in time domain} \label{sec:DopplerDerivation}

In this section, we will derive the expression of the supplementary GW phase induced by both transverse and longitudinal Doppler effects. Our discussion will be somewhat similar to the one presented in~\cite{Bonvin_2017}. We can picture the inner binary system as emitting GW in a "near zone" situated close to its center-of-mass. Then, the lowest-order quadrupole formula gives the plus and cross polarizations of the GW received at the detector as~\cite{Maggiore:1900zz}
\begin{align}\label{eq:hplushcross}
\begin{split}
h_+(t) &= \frac{4}{d} \big( G_N M_c \big)^{5/3} \big( 2 \pi \dot \Phi(t_\mathrm{ret}) \big)^{2/3} \bigg( \frac{1+\cos^2 \iota}{2} \bigg) \cos \Phi(t_\mathrm{ret}) \\
h_\times(t) &= \frac{4}{d} \big( G_N M_c \big)^{5/3} \big( 2 \pi \dot \Phi(t_\mathrm{ret}) \big)^{2/3} \cos \iota \sin \Phi(t_\mathrm{ret})
\end{split}
\end{align} 
where $d$ is the distance to the source, $M_c = (m_1 m_2)^{3/5}/m^{1/5}$ is the chirp mass of the inner binary, $\Phi$ is (twice) its total phase, and $t_\mathrm{ret}$ the retarded time defined by
\begin{equation}\label{eq:tret}
t_\mathrm{ret} = t - d_\mathrm{O,CM}(t_\mathrm{ret})
\end{equation}
where $d_\mathrm{O,CM} = \vert \bm d_\mathrm{O,CM} \vert$ is the norm of the distance of the observer to the center-of-mass of the inner binary. Introducing now the distance from the observer to the \textit{total} center-of-mass of the triple system $\bm d_\mathrm{O,CMT}$ and the line-of-sight vector $\bm n = \bm d_\mathrm{O,CMT} / d_\mathrm{O,CMT}$, one can write $d_\mathrm{O,CM} \simeq d_\mathrm{O,CMT} + \bm n \cdot \bm Y_\mathrm{CM}$ up to negligible corrections in $1/d_\mathrm{O,CMT}$. To simplify the following discussion, we have assumed here that the cosmological redshift and proper motion of the entire system are small; however, using well-known properties of waveforms, one can obtain the GW amplitude and phase at a cosmological redshift $z_0$ simply by replacing the distance $d$ with the luminosity distance $d_L$, multiplying all times, masses and distances by $(1+z_0)$ while frequencies get enhanced by $1/(1+z_0)$.

 We will neglect all Doppler corrections to the amplitude of the GW, focusing only on the phase which is the observable measured with the greatest precision in interferometers (in~\cite{2022arXiv220508550S} it was shown that Doppler amplitude corrections induce a 2\% uncertainty in the estimation of the luminosity distance to the source, which is much smaller than the measurement error for this parameter). To find the phase $\Phi$, one should integrate the GW frequency $f_O$ received by the observer over time. However, due to the motion of the center-of-mass of the inner binary, this frequency differs from the one in the source frame $f_S$ where we can apply standard tools to compute the time-evolution of the frequency \textit{via} an energy flux. More precisely, in the source frame the differential equation governing $f_S$ is~\cite{Maggiore:1900zz}
\begin{equation} \label{eq:dfdts}
\frac{\mathrm{d}f_S}{\mathrm{d}t_S} = \frac{96}{5} \pi^{8/3} \big( G_N M_c \big)^{5/3} f_S^{11/3}\,.
\end{equation}
The redshift factor $z$ relating observer and source frame can be found by expanding the proper time of the source $\mathrm{d}t_S$ :
\begin{equation}
\label{eq: redshift third body}
\frac{1}{1+z} = \frac{\mathrm{d}t_S}{\mathrm{d}t} = \sqrt{-g_{\mu \nu} V_\mathrm{CM}^\mu V_\mathrm{CM}^\nu} \simeq 1 - \frac{V_\mathrm{CM}^2}{2} - \frac{G_N m_3}{R}\,,
\end{equation}
where we have expanded the metric $g_{\mu \nu}$ for small gravitational fields, and $V_\mathrm{CM}^\mu = (1, \bm V_\mathrm{CM})$ is the velocity 4-vector of the center-of-mass of the inner binary. Note that the latter expression contains both a boost factor and the gravitational redshift, the combination of which we loosely denote as "transverse Doppler effect". Once again, note that to keep the discussion as simple as possible we did not display the cosmological redshift factor $z_0$ in Eq.~\eqref{eq: redshift third body}, since it can be taken into account by simply following the prescription described below Eq.~\eqref{eq:tret}.
Using the definition of osculating elements given in~\ref{sec:osculating}, one finds that the redshift $z$ is given by
\begin{equation}\label{eq:z}
z =  \frac{G_N m_3}{a_3} \bigg( \frac{1+X_3}{1-e_3 \cos \eta_3} - \frac{X_3}{2} \bigg)\,,
\end{equation}
where $X_3 = m_3/M$ is the mass ratio of the outer mass $m_3$, and we recall that $\eta_3$ is the outer orbit eccentric anomaly defined by $\eta_3 - e_3 \sin \eta_3 = 2 \pi (t-t_c) / P_3 + \varphi_3$ where $t_c$ is the time at coalescence (so that $\eta_3 < 0$). Integrating equation~\eqref{eq:dfdts} with respect to the time $t$, we find the time-evolution of the source frequency $f_S$:
\begin{equation}\label{eq:fs}
f_S(t) = \frac{5}{8 \pi} \big( 5G_N M_c \big)^{-5/8} \bigg[ \bigg(1-\frac{G_N m_3}{a_3} \bigg(1+\frac{X_3}{2} \bigg) \bigg) \big(t_c -t \big) - \frac{G_N m_3(1+X_3) e_3}{a_3} \frac{P_3}{2 \pi} \big( \sin \eta_3^c - \sin \eta_3 \big) \bigg]^{-3/8}\,,
\end{equation}
where $\eta_3^c$ is the value of the outer eccentric anomaly at coalescence, $\eta_3^c - e_3 \sin \eta_3^c = \varphi_3$. Note that in this simple quadrupolar approximation the frequency diverges at coalescence. Using that $f_O = f_S / (1+z)$, we can now find the phase using $\Phi = 2 \pi \int \mathrm{d}t \; f_O$. Keeping terms only to first order in $G_N m_3 / a_3$ (i.e. to 1PN order for the outer orbit), we find that the integral can be performed analytically and the phase reads
\begin{align} \label{eq:Phi}
\begin{split}
\Phi &= -2 \bigg( \frac{t_c-t}{5 G_N M_c} \bigg)^{5/8} \bigg[ 1 - \frac{5}{16} \frac{G_N m_3}{a_3} \big( 2 +  X_3 \big) \bigg] + \frac{5}{8 \pi} P_3  \big( 5 G_N M_c \big)^{-5/8}  \frac{G_N m_3}{a_3} \big(1+X_3\big) e_3  \frac{\sin \eta_3^c - \sin \eta_3 }{ \big( t_c - t \big)^{3/8}}\,,
\end{split}
\end{align}
where we have chosen to normalize all phase factors to zero at coalescence.
This formula give the transverse Doppler shift due to the motion of the center-of-mass of the inner binary (the longitudinal Doppler effect coming only from the retarded time used in Eq.~\eqref{eq:hplushcross}, see next Section or Ref.~\cite{2019PhRvD..99b4025C}). Let us make a few comments about this equation. The first term in the phase can be identified with a modification of the leading-order PN coefficient (which is proportional to $(t_c-t)^{5/8}$). As such, it would be impossible to determine the parameters of the three-body system using this term only, as they would be completely degenerate with the chirp mass $M_c$; this situation is analoguous to the effect of a constant cosmological redshift. However, note that it is still important to include the corrections proportional to $G_N m_3/a_3$ when doing parameter estimation since it can lead to biases in the measured value of $M_c$. On the other hand, the second term in the phase $\Phi$ presents a more complicated time-dependence. While we will analyze different limits of this transverse Doppler shift in Section~\ref{sec:qualitative}, let us just state here that, contrary to the longitudinal Doppler effect, it does not depend on the product $a_3 \sin \iota_3$ and consequently can potentially allow to break this degeneracy.

\subsection{Stationary Phase Approximation} \label{sec:SPA}

In order to use waveforms in data analysis, it is necessary to compute their Fourier transform. In this section, we will derive the expression of Doppler phase shifts in the Fourier domain, using the stationary phase approximation (SPA) to compute the integral. The Fourier transform of the $+$ polarization reads (analogous formulas hold true for the $\times$ polarization):
\begin{equation}
\tilde h_+(f) = \int \mathrm{d}t \; A(t_\mathrm{ret}) \cos \Phi(t_\mathrm{ret}) e^{2i \pi f t} \; , \quad A(t_\mathrm{ret}) =  \frac{4}{d} \big( G_N M_c \big)^{5/3} \big( 2 \pi \dot \Phi(t_\mathrm{ret}) \big)^{2/3} \bigg( \frac{1+\cos^2 \iota}{2} \bigg)\,.
\end{equation}
We switch to the retarded time as an integration variable. As stated previously, we will neglect all Doppler corrections to the amplitude of the GW, keeping only the phase factors. Thus, we can assume that the Jacobian of the change of variable is $\mathrm{d}t_\mathrm{ret}/\mathrm{d}t = 1$ and we have
\begin{equation}
\tilde h_+(f) = e^{2 i \pi f d_\mathrm{O,CMT}} \int \mathrm{d}t_\mathrm{ret} \;  A(t_\mathrm{ret}) \cos \Phi(t_\mathrm{ret}) e^{2i \pi f ( t_\mathrm{ret} + \bm n \cdot \bm Y_\mathrm{CM})}\,,
\end{equation}
where $d_\mathrm{O,CMT}$ and $\bm n$ were introduced below Eq.~\eqref{eq:tret}. Splitting $\cos \Phi = (e^{i \Phi} + e^{-i \Phi})/2$, we see that there is a stationary point at the time $t^*(f)$ defined by
\begin{equation} \label{eq:tstar}
2 \pi f = \left. \frac{\dot \Phi}{1+ \bm n \cdot \bm V_\mathrm{CM}} \right\vert_{t^*(f)}\,.
\end{equation}
Evaluating the integral by expanding the integrand to second order around the stationary point $t^*(f)$, and neglecting once again all Doppler corrections to the amplitude, we find that $\tilde h_+(f) = \mathcal{A} e^{i \Psi}$ with
\begin{equation} \label{eq:amplPhase}
\mathcal{A} = \frac{1}{\pi^{2/3}} \bigg( \frac{5}{24} \bigg)^{1/2} \frac{1}{d} \big( G_N M_c \big)^{5/6} \frac{1}{f^{7/6}} \bigg( \frac{1+\cos^2 \iota}{2} \bigg) \; , \quad \Psi = 2 \pi f \big( t^* + d_\mathrm{O,CMT} + \bm n \cdot \bm Y_\mathrm{CM}(t^*) \big) - \Phi(t^*) - \frac{\pi}{4}\,.
\end{equation}
Remains to solve Eq.~\eqref{eq:tstar} defining $t^*(f)$ and plug it back in the phase. This can be done perturbatively since $V_\mathrm{CM} \ll 1$. Since $\dot \Phi = f_S (1-z)$ where $f_S$ and $z$ are given in Eqs.~\eqref{eq:fs} and~\eqref{eq:z}, there are also corrections of order $V_\mathrm{CM}^2$ in Eq.~\eqref{eq:tstar} defining $t^*$, coming from the transverse Doppler effect. We thus split $t^* = t_{(0)}^*+t_{(1)}^*+t_{(2)}^*$ where $t_{(1)}^*$ is of order $V_\mathrm{CM}$ and $t_{(2)}^*$ of order $V_\mathrm{CM}^2$, while $t_{(0)}^*$ is the usual time to coalescence in the quadrupolar approximation:
\begin{equation}
t_c - t_{(0)}^*(f) = \frac{5}{256} \big( G_N M_c \big)^{-5/3} \big( \pi f \big)^{-8/3}
\end{equation}
this expression being found by equating $f$ to the lowest-order $f_S$ shown in Eq.~\eqref{eq:fs}. Then, by perturbatively solving Eq.~\eqref{eq:tstar} one finds
\begin{equation}\label{eq:t1star}
t_{(1)}^* = \frac{5}{96} \bm n \cdot \bm V_\mathrm{CM} \big( G_N M_c \big)^{-5/3} \big( \pi f \big)^{-8/3}\,,
\end{equation}
while the explicit expression of $t_{(2)}^*$ will not be needed in the following. We now expand the phase defined in Eq.~\eqref{eq:amplPhase} up to second order in $V_\mathrm{CM}$ (i.e., to the order at which the transverse Doppler effect shows up). Using that at lowest order $\dot \Phi = 2 \pi f$ we find that indeed the term containing $t_{(2)}^*$ simplifies from the phase and using Eqs.~\eqref{eq:Phi} and~\eqref{eq:t1star} we are left with
\begin{align}\label{eq:FourierSpaceDoppler}
\begin{split}
\Psi &= \Psi_0 + 2 \pi f \bm n \cdot \bm Y_\mathrm{CM}  + \frac{1}{32} \big(\pi G_N M_c f)^{-5/3} \bigg[  \frac{5}{3} \big( \bm n \cdot \bm V_\mathrm{CM} \big)^2 - \frac{5}{8} \frac{G_N m_3}{a_3} \big( 2+  X_3 \big) \bigg] \\
  &+ P_3 f \frac{G_N m_3}{a_3} \big(1+X_3\big) e_3  \sin \eta_3 \,,
\end{split}
\end{align}
where $\Psi_0 = 2 \pi f t_c' - \frac{\pi}{4} + 3 \big(\pi G_N M_c f)^{-5/3} / 128$ is the usual phase of an isolated binary system in the quadrupole approximation, and we have absorbed an unimportant constant in the definition of the time at coalescence $t_c'$. Note that $\eta_3$ implicitly depends on $f$ through its definition $\eta_3 - e_3 \sin \eta_3 = 2 \pi (t_{(0)}^*(f)-t_c)/P_3 + \varphi_3$.
In the above formula~\eqref{eq:FourierSpaceDoppler}, the first non-trivial term $\Psi_{||} = 2 \pi f \bm n \cdot \bm Y_\mathrm{CM}$ is the lowest-order longitudinal Doppler shift which has been discussed at length in the literature~\cite{Robson_2018, 2019MNRAS.488.5665W, Randall_2019, Bonvin_2017, Inayoshi:2017hgw, https://doi.org/10.48550/arxiv.2109.08154, Toubiana_2021, 2022arXiv220508550S, 2019PhRvD..99b4025C}. The other terms are suppressed by $V_\mathrm{CM}$ with respect to this lowest-order Doppler shift and contain both longitudinal and transverse components. Note that, as discussed below Eq.~\eqref{eq:Phi}, the terms proportional to a constant multiplying $\big(\pi G_N M_c f)^{-5/3}$ just renormalize the measured value of the chirp mass, and as such do not permit a measurement of the parameters of the three-body system. Thus, the only useful transverse Doppler term for data analysis purposes is the one on the second line that we denote by $\Psi_\perp$. In terms of the osculating elements introduced in Section~\ref{sec:osculating}, the scalar products $\bm n \cdot \bm Y_\mathrm{CM}$ and $\bm n \cdot \bm V_\mathrm{CM}$ read
\begin{align}
\bm n \cdot \bm Y_\mathrm{CM} &= - X_3 a_3 \sin \iota_3 \big[ \big( \cos \eta_3-e_3\big) \sin \omega_3 + \sqrt{1-e_3^2} \sin \eta_3 \cos \omega_3  \big]\,, \label{eq:nDotYCM}  \\
\bm n \cdot \bm V_\mathrm{CM} &= - X_3 \sqrt{\frac{G_N M}{a_3}} \frac{\sin \iota_3}{1-e_3 \cos \eta_3} \big[  \sqrt{1-e_3^2} \cos \eta_3 \cos \omega_3 - \sin \eta_3 \sin \omega_3  \big]\,. \label{eq:nDotVCM}
\end{align}


Before moving on and analyzing some qualitative properties of the Fourier space Doppler shift in~\eqref{eq:FourierSpaceDoppler}, let us make a comment about the validity of the perturbative expansion that we used to evaluate $t^*$. In our derivation, there is an implicit assumption about the fact that the time-to-frequency map implied by Eq.~\eqref{eq:tstar} is singled-valued. However, if the outer perturber $m_3$ is really close to the inner binary, this assumption could be violated (see e.g.~\cite{Toubiana_2021, 2022arXiv220508550S} for an interesting discussion). This phenomenon can happen when $\mathrm{d}f/\mathrm{d}t \leq 0$. This means that the Doppler shift due to the acceleration of the outer binary, $2 \pi f \bm n \cdot \bm A_\mathrm{CM}$, has to be greater than the usual chirping due to radiation-reaction given in Eq.~\eqref{eq:dfdts}. Consequently, our computations are valid only when $P_3$ is larger than a limiting value $P_3^\mathrm{lim}$:
\begin{equation}
\label{eq: limit P3}
P_3 \geq P_3^\mathrm{lim} = 11 \mathrm{yr} \; \big( \sin \iota_3 \big)^{3/4} \bigg(\frac{f }{0.01 \mathrm{Hz} }\bigg)^{-2} \bigg( \frac{M_c}{30 \msun} \bigg)^{-5/4} \bigg( \frac{m_3}{4 \times 10^6 \msun} \bigg)^{1/4}\,.
\end{equation}
For such small outer periods, we would anyway expect that the waveform is impacted by more relativistic effects than the ones that we considered in this article, like e.g. the Shapiro time delay discussed in~\cite{2022arXiv220508550S}. Thus, we will only consider outer periods greater than $P_3^\mathrm{lim}$ in all our analysis.

Finally, note that as underlined below Eq.~\eqref{eq:tret}, all quantities appearing in the phase~\eqref{eq:FourierSpaceDoppler} should be understood as \textit{detector frame} quantities, related to source frame ones by a cosmological redshift factor $(1+z_0)$: $P_3 = (1+z_0) P_3^\mathrm{s}$, $a_3 = (1+z_0) a_3^\mathrm{s}$, $m_3 = (1+z_0) m_3^\mathrm{s}$, $M_c = (1+z_0) M_c^\mathrm{s}$, $f = f^\mathrm{s}/(1+z_0)$ where the s superscript denotes a quantity in source frame.


\section{Qualitative analysis: limits and degeneracies} \label{sec:qualitative}

In this section we will analyze the preceding formula~\eqref{eq:FourierSpaceDoppler} for the Doppler phase shift in different limits and discuss the measurability of the parameters of the outer orbit as well as their degeneracies. In the following, we assume that we know the time at coalescence $t_c$ from an analysis of the waveform of the inner binary; otherwise we can absorb the outer phase $\varphi_3$ in a shift of the initial time.

\subsection{Generic case}

In the generic case, the question of whether there exist degeneracies among parameters of the three-body system is essentially the same than asking what is the frequency-dependence of each term in the phase~\eqref{eq:FourierSpaceDoppler}. Indeed, if two terms feature the same frequency-dependence, then their amplitude cannot be determined separately and we are in the presence of a degeneracy. In the phase~\eqref{eq:FourierSpaceDoppler}, the only transverse Doppler term useful for data analysis purposes is $\Psi_\perp$ in the second line, since the transverse Doppler component of the first line is degenerate with the chirp mass as discussed above. However, the frequency-dependence of this term is completely degenerate with the lowest-order longitudinal Doppler term, cf Eq.~\eqref{eq:nDotYCM}. Letting aside the second-order longitudinal Doppler shift $(\bm n \cdot \bm V_\mathrm{CM})^2$ for the moment, we are thus led to ask how many parameters of the outer orbit can be measured with the first-order longitudinal Doppler term.

As already emphasized below Eq.~\eqref{eq:FourierSpaceDoppler}, the eccentric anomaly depends on frequency through $\eta_3 - e_3 \sin \eta_3 = 2 \pi (t_{(0)}^*(f)-t_c)/P_3 + \varphi_3$. Thus, the frequency-dependence of the eccentric anomaly is strongly impacted by the parameters $e_3$, $P_3$ and $\varphi_3$, and we can hope for a good determination of these three parameters. On the other hand, the $\bm n \cdot \bm Y_\mathrm{CM}$ term depends on the eccentric anomaly only through two terms $\cos \eta_3$ and $\sin \eta_3$, while there remains three parameters which multiply the amplitude of these terms, $a_3$, $\iota_3$ and $\omega_3$. Thus, the lowest-order longitudinal Doppler effect alone with the transverse Doppler shift would not enable to measure all parameters of the three-body system. Fortunately, the second-order longitudinal Doppler term $(\bm n \cdot \bm V_\mathrm{CM})^2$ saves the day, since it features additional dependence on $\eta_3$ like e.g. $\cos 2 \eta_3$. Thus, in the generic case we can \textit{a priori} hope to measure all parameters of the three-body system without any remaining degeneracy. We will now examine different limits of the phase shift~\eqref{eq:FourierSpaceDoppler}.

\subsection{Small eccentricity} \label{sec:small_e}

In the small-eccentricity limit $e_3 \rightarrow 0$, the transverse Doppler term on the second line of Eq.~\eqref{eq:FourierSpaceDoppler} vanishes and there remains only longitudinal Doppler terms. In this case, one can show that there is an exact degeneracy $a_3 \sin \iota_3$ which we mentioned in the Introduction. This is evident from the expression of $\bm n \cdot \bm Y_\mathrm{CM}$ given in Eq.~\eqref{eq:nDotYCM}, while for the $\bm n \cdot \bm V_\mathrm{CM}$ one has to use Kepler's law to trade the mass $M$ for the period $P_3$ used in data analysis: $(G_N M/a_3)^{1/2} = 2 \pi a_3 / P_3$. Furthermore, it turns out that only the combination $\varphi_3 + \omega_3$ enters the scalar products~\eqref{eq:nDotYCM} and~\eqref{eq:nDotVCM}. Thus, in this case one can measure only three parameters of the outer orbit, $P_3$, $a_3 \sin \iota_3$ and $\varphi_3 + \omega_3$, which are the parameters discussed in previous studies using only the longitudinal Doppler shift~\cite{Robson_2018, 2019MNRAS.488.5665W, Randall_2019}.

\subsection{Face-on systems}

When the system is observed face-on $\iota_3 = 0$, the longitudinal Doppler shift vanishes. Since the transverse Doppler term does not depend on $\omega_3$, one can measure the parameters $P_3$, $e_3$ and $\varphi_3$ from the frequency-dependence of this term, and finally the parameter $a_3$ from the overall amplitude. Thus, this completely determines the parameters which are most relevant for astrophysics.

\subsection{Large outer period}\label{sec:gammas}

The qualitative frequency-dependence of the Doppler phase shift~\eqref{eq:FourierSpaceDoppler} will mainly depend on the ratio between the outer period $P_3$ and the observation time $T_\mathrm{obs}$ (which we consider to be of the order of the radiation-reaction timescale for BHs systems which have a non-negligible chirping in the LISA band). Indeed, if $P_3 < T_\mathrm{obs}$ we will observe more than one revolution of the outer orbit during the observation time; consequently, we will be able to measure the amplitude of terms oscillating with the outer period like $2 \pi f \bm n \cdot \bm Y_\mathrm{CM}$. Since the frequency-dependence of all Doppler terms is quite different, one can hope for a good determination of the parameters of the outer orbit in this case.
On the other hand, if $P_3 \gg T_\mathrm{obs}$ we probe only a small portion of the outer orbit. In this case, one can expand all Doppler terms for $\eta_3^c-\eta_3 \ll 1$ and one generically finds that the frequency-dependence of the phase is, up to an irrelevant constant shift of the time at coalescence,
\begin{equation} \label{eq:def_gammas}
\Psi \simeq \gamma_1 \big( \pi G_N M_c f \big)^{-5/3} + \gamma_2  \big( \pi G_N M_c f \big)^{-13/3}  + \gamma_3  \big( \pi G_N M_c f \big)^{-7} + \dots
\end{equation}
where the $\gamma$'s are constants depending on the parameters of the outer orbit, with the limit that $\gamma_1 = 3/128$, $\gamma_2 = \gamma_3 = \dots =0$ when $a_3$ is sent to infinity (or equivalently when $m_3 = 0$). More precisely, dimensional analysis shows that the scaling of $\gamma_n$ for $n \geq 2$ is
\begin{equation}\label{eq:scalingGammas}
\gamma_n \sim \bigg( \frac{G_N M_c}{P_3} \bigg)^{n-1} \times \bigg( \frac{G_N m_3}{a_3} \bigg)^{k}\,,
\end{equation}
where $k=1/2$ for the lowest-order longitudinal Doppler effect, and $k=1$ for transverse Doppler.
Thus, the Doppler shift is encapsulated in a series of post-Newtonian coefficients, $\gamma_1$ being a change in the leading-order quadrupole phase, $\gamma_2$ corresponding to a -4PN term, $\gamma_3$ to -8PN, and so on and so forth (this scaling can be seen from the fact that $\big( G_N M_c f \big)^{2/3}$ is the traditional 1PN order frequency parameter for the inner binary). Of course, even if they appear at negative PN orders, these terms are more and more suppressed by the ratio $G_N M_c/P_3$ which is small for large outer period, so that the amplitude of these terms is smaller and smaller. Eventually, one will reach a point in the expansion where the term parametrized by $\gamma_n$ gives a contribution to the phase smaller than the observability criterion.

Thus, depending on the actual value of the ratio $P_3/T_\mathrm{obs}$ and the SNR, only a finite number of PN coefficients can be measured and this can be insufficient to determine all the parameters of the outer orbit. For example, if only $\gamma_1$ and $\gamma_2$ can be measured (as is discussed e.g. in~\cite{Toubiana_2021}), then since $\gamma_1$ is used to determine the chirp mass one is left with one measured parameter containing all the 6 parameters characterizing the outer orbit. In this case, it makes more sense to measure only the amplitude of this -4PN coefficient without trying to recover the parameters of the outer orbit, as has been proposed in~\cite{Toubiana_2021}, where it is also shown that this -4PN coefficient is degenerate with other environmental effects.

\subsection{Observability criterion in the $(m_3, P_3)$ plane}

\begin{figure}
\centering
\includegraphics[width=0.7\columnwidth]{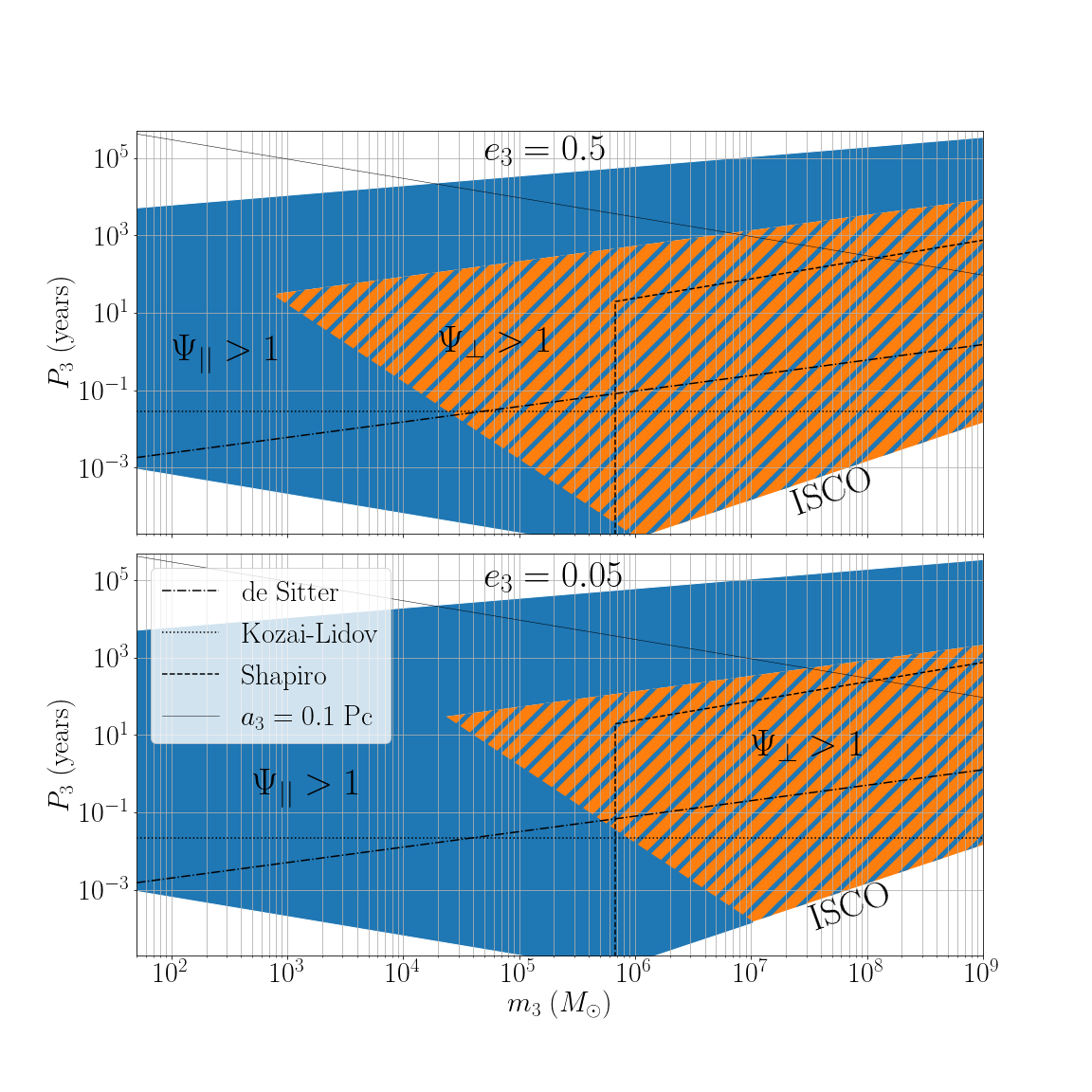}
\caption{Observability of longitudinal and transverse Doppler effects in the $(m_3,P_3)$ plane. The Doppler phase in Eq.~\eqref{eq:FourierSpaceDoppler} is greater than 1 radian in the colored area: in blue the longitudinal Doppler phase $\Psi_{||} > 1$ and in orange the transverse Doppler phase $\Psi_\perp > 1$, where $\Psi_{||}$ and $\Psi_\perp$ were introduced below Eq.~\eqref{eq:FourierSpaceDoppler}. 
In the upper plot, we take the parameters of the inner binary to be $M_c = 30 \msun$ and $f=12$mHz, while for the outer binary we fix $e_3 = 0.5$, $\iota_3 = \omega_3 = \varphi_3 = \pi/4$. The lower plot has $e_3=0.05$ and same other parameters.
Also shown is the Innermost Stable Circular Orbit (ISCO) of the outer orbit, so that no triple system can live on the bottom right corner of the plot. Furthermore, we also plot the lines where other effects (Shapiro time delay, de Sitter precession, Kozai-Lidov oscillations) give a measurable phase shift: below these lines, these effects are observable in principle. Note that for high masses, the transverse Doppler effect covers the largest portion of parameter space among all degeneracy-breaking effects even when the eccentricity $e_3$ is small. Finally, we also show the period corresponding to a semimajor axis $a_3=0.1$ Pc, below which it has been suggested that the detection rate of binaries produced by the interaction channel around SMBH could be as high as $10-100$ events per year in advanced LIGO~\cite{O_Leary_2009, Antonini_2016_2, Petrovich_2017, VanLandingham_2016}.
}
\label{fig:ParamSpace}
\end{figure}

Finally, one can ask the question of what is the portion of parameter space where the phase shift given in Eq.~\eqref{eq:FourierSpaceDoppler} is observable. While we will answer more quantitatively to this question in Section~\ref{sec:MCMC}, here we will just give some useful order-of-magnitudes estimates by requiring that the phase shift induced by Doppler terms is greater than 1 radian. When the outer period is large, this is equivalent to require than $\gamma_2  \big( \pi G_N M_c f \big)^{-13/3} \geq 1$, where $\gamma_2$ has been introduced in the last Section~\ref{sec:gammas} (the term proportional to $\gamma_1$ is completely degenerate with the chirp mass and cannot be used to detect the Doppler effect). From the scaling of $\gamma_2$ shown in Eq.~\eqref{eq:scalingGammas}, we easily see that it imposes a maximal value of $P_3$ above which Doppler terms are unobservable.

On the other hand, when the outer period $P_3$ is of the order or less than the observation time, the requirement of observability translates in a \textit{minimal} bound on $P_3$ in order for Doppler effects to be observable. This can be seen from the fact that the amplitude of terms oscillating with $\eta_3$ in Eq.~\eqref{eq:FourierSpaceDoppler} is proportional to $a_3 \sim (G_N M P_3^2)^{1/3}$ (for the $\bm n \cdot \bm Y_\mathrm{CM}$ term) or $P_3 G_N m_3/a_3 \sim (G_N^2 M^2 P_3)^{1/3}$ (for the transverse Doppler term on the second line). Thus, there is actually a bounded zone in the $(m_3, P_3)$ plane where longitudinal or transverse Doppler terms are observable. This section of parameter space is shown in Figure~\ref{fig:ParamSpace}.

In the same Figure~\ref{fig:ParamSpace}, we have also shown for illustrative purposes the sections of parameter space where other effects already discussed in the existing literature give a phase shift greater than 1, allowing to further break the degeneracies among parameters. These effects are: the Shapiro time delay~\cite{2022arXiv220508550S}, the de Sitter precession of the BBH angular momentum~\cite{Yu_2021}, and the Kozai-Lidov oscillations~\cite{Chandramouli:2021kts}. To estimate the phase shift due to the Shapiro time delay, we use the formula given in~\cite{2022arXiv220508550S}, separating the two limiting cases of a period greater or smaller than the observation time as before. However, concerning the Kozai-Lidov and de Sitter precession effects, we simply use the criterion presented in~\cite{Yu_2021}, stating that the period of these effects should be less than $\sim100$ yr to allow for a detection by a network of interferometers.
Overall, we see that the transverse Doppler effect covers the largest portion of parameter space among all degeneracy-breaking effects, particularly in the case where the outer perturber is a SMBH which we will consider in the following.

Finally, we also illustrate in Figure~\ref{fig:ParamSpace} how reducing the outer eccentricity $e_3$ affects the magnitude of the transverse Doppler terms. Indeed, we have shown in Section~\ref{sec:small_e} that the transverse Doppler effect vanishes if the outer eccentricity is $e_3=0$. Furthermore,
in the migration traps of AGNs where there could be binaries with strong three-body effects induced by the SMBH, we expect small outer eccentricities as migrating bodies in AGN disks, similarly to planets in protoplanetary disks, are expected to circularize~\cite{2022arXiv220508550S}. However, as one can see in the lower panel of Figure~\ref{fig:ParamSpace}, even for an outer eccentricity as low as $e_3=0.05$ the transverse Doppler effect covers the largest region of parameter space among all degeneracy-breaking effects when the third mass is a supermassive black hole.



\section{Parameter estimation with MCMC} \label{sec:MCMC}

\subsection{Analysis method}

We want to infer the parameters of the triple system, the set of which we denote by $\theta$. The posterior distribution of $\theta$ given the gravitational wave data is denoted as $p(\theta|d)$. Then, Bayes' theorem relates the likelihood $p(d|\theta)$ to the posterior as $p(\theta|d) = p(d|\theta)p(\theta)/p(d)$, where $p(\theta)$ is the prior on $\theta$, and $p(d)$ is a normalization constant (that depends on the data), which is of no interest to us in this particular case. 
We assume that the noise in the interferometer is stationary and Gaussian. Thus, it can be described by the (Fourier transformed) \textit{power spectral density} (PSD) $S_n(f)$. Under this hypothesis, the likelihood is given by~\cite{Maggiore:1900zz} 
\begin{equation}
    p(d|\theta) = \exp\left[-\frac{1}{2}\left(d-h(\theta)|d-h(\theta)\right)\right]\,,
\end{equation}
where $d$ is the data and $h(\theta)$ is the gravitational wave template.
We have defined the scalar product (or overlap) over frequency as 
\begin{equation} \label{eq:overlap}
    \left(d_1|d_2\right) = 4\mathcal{R}\left[\int_0^\infty\frac{d_1(f) d_2(f)^*}{S_n(f)}\mathrm{d}f\right]\,.
\end{equation}
The real part is written as $\mathcal{R}$ and an asterisk indicates the complex conjugate. In fact, for LISA data analysis the overlap shown in Eq.~\eqref{eq:overlap} is a sum of three terms, one for each LISA time-delay interferometry (TDI) observable as described in~\cite{Toubiana_2020}.
In the following simulations of data, we assume that the particular signal has no noise component, that is $d = h(\hat\theta)$, where the hatted quantities represent the true parameters. 
Therefore, the resulting likelihood distributions should peak around the injected values, contrary to what one expects with a non-zero noise contribution to the signal.


Although the use of a Fisher matrix approach to evaluate the uncertainties in estimating the parameters of a LISA source is quite common in the literature (see e.g.~\cite{https://doi.org/10.48550/arxiv.2109.08154,Inayoshi:2017hgw,Randall_2019,2019MNRAS.488.5665W,Robson_2018}), we chose not to compute it and instead evaluate uncertainties by running several MCMCs to estimate the posterior distribution $p(\theta|d)$. The main reason for this choice is that, on top of being inaccurate for sources with low signal-to-noise ratio~\cite{Vallisneri_2008}, the Fisher matrix $\Gamma$ for our three-body system parameter estimation turns out to be very ill-conditioned so that computing its inverse $\Gamma^{-1}$ (representing the uncertainties in parameters) is prone to large numerical uncertainties. This can be understood as follows: the condition number $\kappa$ is approximately the ratio of the largest to the lowest eigenvalue of $\Gamma$, which means that any numerical uncertainty of the order of $1/\kappa$ in the computation of $\Gamma$ will translate in an order-one error in the computation of the inverse $\Gamma^{-1}$. Now, it turns out that there is a large hierarchy in the eigenvalues of $\Gamma$, which is related to the fact that the observable parameters in the Doppler phase for $P_3>T_\mathrm{obs}$ are the constants $\gamma_n$ defined in Eq.~\eqref{eq:def_gammas}: since for larger $n$, $\gamma_n$ leads to a smaller phase shift, its uncertainty will be larger. Typically, we find that for a $m_3 = 10^8 \msun$ central BH and an outer period $P_3 > 40$~yr, the condition number will be $\kappa > 10^8$. Since the integral defining the gravitational-wave overlap entering into the elements of $\Gamma$ is computed with a $10^{-8}$ accuracy, this means that we can not rely on a Fisher-based approach for periods $P_3 > 40$~yr, while we will explore parameter estimation up to $P_3 \lesssim 150$~yr with our MCMC-based approach. 
As we will see in  section \ref{subsec: results}, the approximation of Gaussian errors on the triple parameters is often insufficient to describe the full posterior in the case of periods larger than 90 years.

\subsection{Technical aspects of the MCMC analysis}

We use \texttt{lisabeta} \cite{Marsat:2018oam} to estimate the posterior density distribution for the three-body signal. The \texttt{IMRPhenomD} waveform approximant \cite{Thompson:2020nei, Pratten:2020ceb, Randall_2019} is assumed to describe the observed signal, considering only the dominant $(2,2)$ harmonic. We model the phase shift induced by the three-body system by multiplying the waveform by phase factor $e^{i \Psi}$ in Fourier space, where $\Psi$ is given in Eq.~\eqref{eq:FourierSpaceDoppler}. Note that in particular the dephasing is given in terms of simple functions of the parameters of the three-body system, so that it is very quick to compute numerically and can be efficiently used in the MCMC analysis. This is to be contrasted with the approach adopted in e.g.~\cite{Robson_2018, https://doi.org/10.48550/arxiv.2109.08154, Randall_2019}, where the phase shift depends on an integral and so could be more time-consuming to implement numerically.
For the LISA noise curve, we use the ``SciRDv1'' model \cite{SciRD1}.

We use 200 walkers and -- depending on the of the period of the inner binary -- these undergo 80.000 to 160.000 convergence steps. At least 10.000 steps have to be discarded to make sure the chains are ``burned-in'' (refer to appendix~\ref{app: convergence} for the computation of the Gelman-Rubin criterion as a verification of the chains' convergence). 
The choice of a basis of parameters to sample the probability distribution is of great importance, since a better parametrization can lead to a much faster convergence time of the MCMC. This can be understood from the fact that, in our system, some parameters are measured with a much better precision than other ones. If we were to parametrize the system without identifying the precisely measured parameters, it would take a lot of steps to reveal in parameter space the subtle degeneracies showing the tight constraints on some combination of parameters. On the other hand, using a basis taking into account these degeneracies, the MCMC converges quickly on the parameters with a small variance and spends most of its time only exploring the basis of parameters with a larger variance.

To make this abstract discussion more concrete, let us specialize it to our particular case. We know that the parameters of the three-body system which are measured with the best precision are the ones probed by the longitudinal Doppler shift, with is the largest phase shift caused by the third body on the waveform. As emphasised many times in this article, when using the longitudinal Doppler shift only to estimate the parameters of the three-body system one is faced with degeneracies among them, in particular between the outer mass $m_3$ and the inclination $\iota_3$. Indeed, the amplitude of the longitudinal Doppler shift shown in Eq.~\eqref{eq:FourierSpaceDoppler} is proportional to $a_3 \sin \iota_3 \sim (G_N m_3 P_3^2)^{1/3} \sin \iota_3$, where we have used Kepler's law to obtain this scaling (and the fact that $m_3 \sim M = m_1+m_2+m_3$ in the physical situations considered in this article). Thus, the precisely measured parameter is $m_3^{1/3} \sin \iota_3$. On the other hand, the only way to measure $m_3$ and $\iota_3$ independently is to use the transverse Doppler shift, which is an effect of smaller magnitude so that these parameters will be measured with less precision. We could use the basis $(m_3, \iota_3)$ in our MCMC analysis but because of the aforementioned reason, the convergence would be slow. Instead, we will use the basis $(m_3 \sin^3 \iota_3, m_3)$ since we expect convergence to be very quick on the first parameter.
Actually, we will use for the parameters characterizing the outer orbit the complete basis $(m_3 \sin^3(\iota_3) \sin^3(\varphi_3 + \omega_3), m_3 \sin^3 \iota_3, m_3, P_3, e_3, \varphi_3-\omega_3 )$ since we observe even better convergence results with these parameters. The reason for using this basis is, when taking the small-eccentricity $e_3 \rightarrow 0$ and large period $P_3 \gg T_\mathrm{obs}$ limits (which is an approximate limit of the parameters we use), one can observe that the lowest-order longitudinal Doppler effect is proportional to $m_3^{1/3} \sin \iota_3 \sin(\varphi_3+\omega_3)$. When displaying the results of the analysis, we will finally convert the probability distribution to the "physical space" of parameters $(m_3, P_3, e_3, \iota_3, \omega_3, \varphi_3)$.

We will assume that the binary is observed as it inspirals through the LISA sensitivity band in frequency, and then at a later stage it is observed by a ground based detector. Thus, for the present analysis, we assume that all inner binary parameters such as chirp mass, mass ratio, inclination, sky position, etc.\,are measured with such high precision that we assume them to be known.
(Appendix~\ref{app: impact fixing chirp mass} quantifies the impact from fixing the binary's chirp mass on the uncertainty on $m_3$.)
When the inner binary parameters are not fixed, the parameters that describe the outer orbit are much less constrained -- see the discussion in Ref.~\cite{2022arXiv220508550S}. This loss of precision is due to the correlations among inner and outer binary parameters. For example, in the case of large outer periods discussed in Section~\ref{sec:gammas}, the time at coalescence and the chirp mass are degenerate with some combination of the parameters of the three-body system. 
For the priors on the triple system parameters applied, please consider appendix \ref{app: priors}. Note that in order to accelerate convergence, we use a narrow range of priors around the true value of the system parameters. This is only used for convenience, and by no means we pretend to provide here a realistic data analysis of LISA. The full challenge of parameter estimation of a three-body system given a strain data with unknown signal is a very complicated task which is way beyond the scope of this paper. Still, it would be an important avenue for future work. It is also interesting to note that some detections of triple systems could be missed in LISA if we were to use only vacuum two-body templates, since the waveform would not be accurate enough for a matched-filter analysis (see the more in-depth discussion in~\cite{2022arXiv220508550S}). 

\subsection{Results}
\label{subsec: results}

\begin{table}[h!]
\renewcommand{\arraystretch}{1.3} 
\begin{center}
\begin{tabular}{||c c c||} 
 \hline
 Parameter & Value & Dimension\\ [0.5ex] 
 \hline\hline
 $M$ &70 & M$_\odot$ \\
 $q$ & 1.3& - \\
 $d_L$ &250 & Mpc \\
 $\iota$ &0.5& -\\
 $\beta$ &1.0472& -\\
 $\lambda$ &1.9& -\\
 $\chi_1$ &0&- \\
 $\chi_2$ &0&- \\
 $\phi$ &0.7&- \\
 $\psi$ &1.2&- \\
 $f_{\rm start}$ &0.01272& Hz \\
 \hline
 $m_3$ &$10^7$/$10^8$/$10^9$ & M$_\odot$ \\
 $\iota_3$ &0.8& -\\
 $\omega_3$&0.35& -\\
 $P_3$ &varying& yr\\
 $e_3$ &0.1& -\\
 $\phi_3$ &0.7&- \\
 \hline
\end{tabular}
\caption{Parameters of the hierarchical system considered. The first set of parameters characterizes the inner binary, while the second describes the outer orbit and are associated to the Doppler shift. Both sets are defined in the main text below Eq.~\eqref{eq:CM}. Note that the masses are given in the detector frame.  }
\label{tab:parameters triple}
\end{center}
\end{table}

\begin{figure}
\includegraphics[width=0.7\columnwidth]{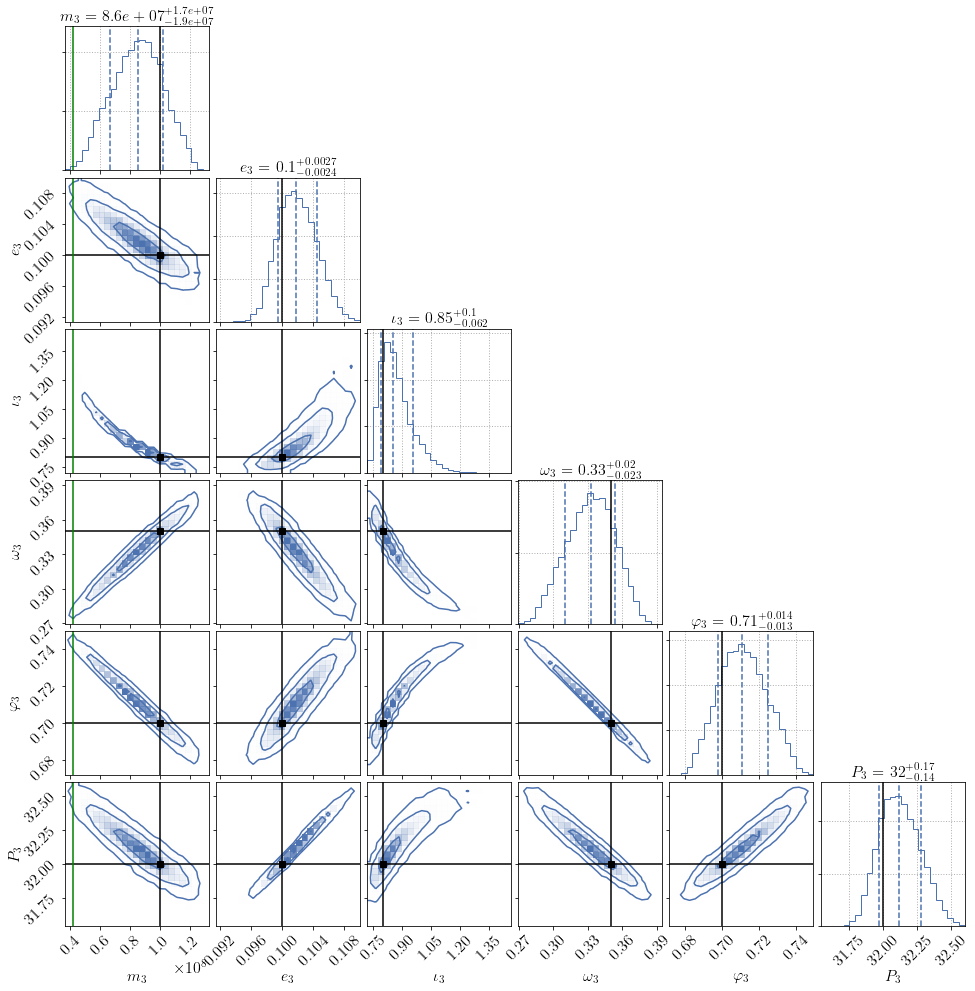}
\caption{The result of the MCMC analysis for a SOBH that is in orbit around a SMBH with $m_3=10^8\msun$, and an eccentricity $e_3=0.1$. The true parameters are indicated with a black line. We have converted the posterior samples from the unphysical parameters $m_3 \sin^3(\iota_3) \sin^3(\varphi_3 + \omega_3)$, $m_3 \sin^3(\iota_3)$ and $\varphi_3 - \omega_3$ back to the variables of interest $\iota_3,\omega_3$ and $\varphi_3$. However, in the figure it remains clearly visible that a combination of $m_3$ and $\iota_3$ is much better measured than the parameters in themselves, as well as some combinations of angles.
One can measure the mass of the central BH with a precision of $\sim 20\%$. Note also that the initial phase $\varphi$ is sampled over in the interval $\left[0,\pi\right]$, but not shown here.
We have also added as a green line the minimum mass $m_3^\mathrm{min}$ corresponding to the parameter which we would estimate using the longitudinal Doppler shift only (see discussion in the main text).}
\label{fig:example result 10^8}
\end{figure}

\begin{figure}
\includegraphics[width=0.63\columnwidth]{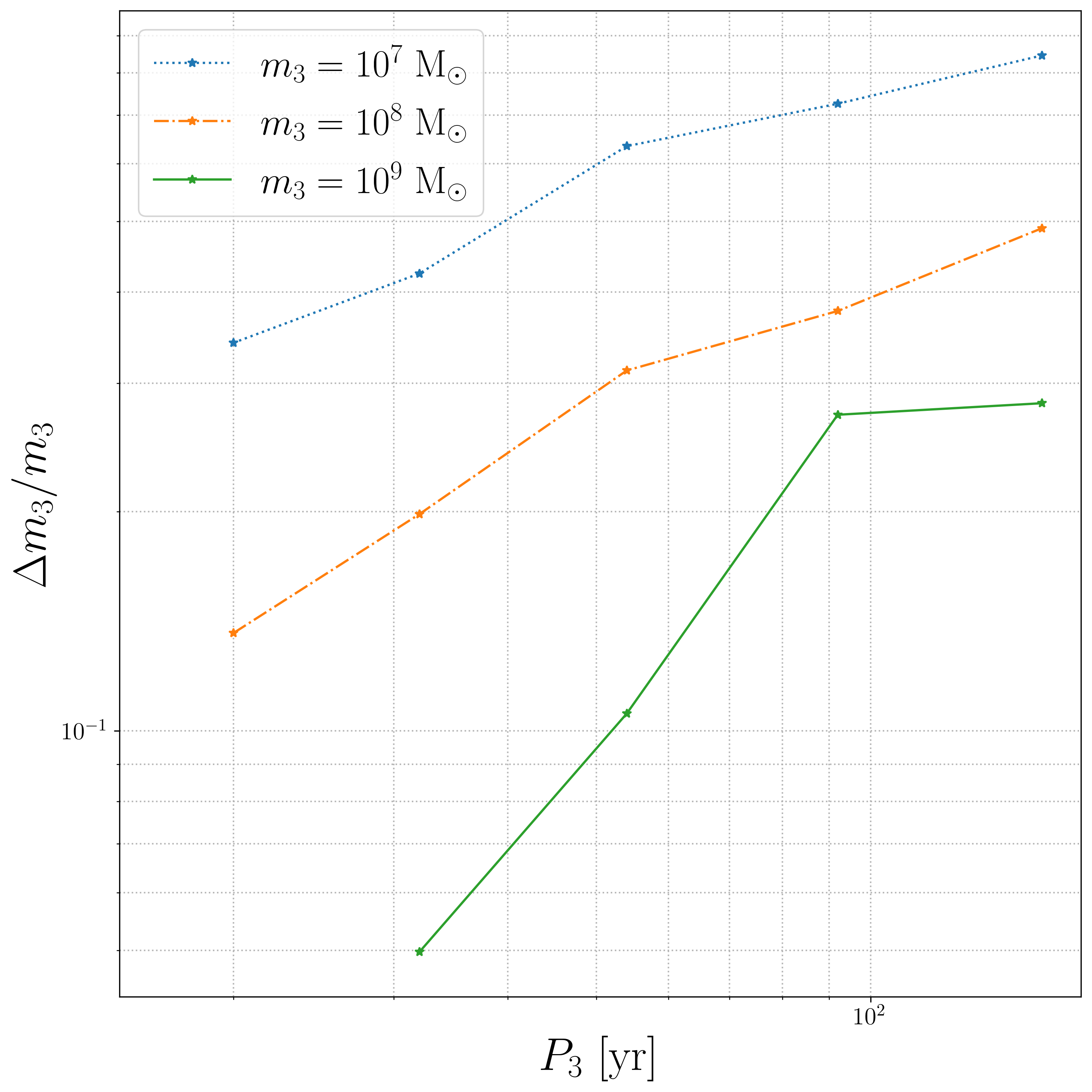}
\caption{The relative error on the mass of the third body $m_3$ as a function of the period of the outer binary $P_3$.
This is computed for three different values of $m_3$. In case of the central mass of $m_3 = 10^8\,\msun$ ($m_3 = 10^7\,\msun$ and $m_3 = 10^9\,\msun$) the results are drawn in orange (blue and green). For a mass of $m_3 = 10^9 \,\msun$, we omit the binary with a period of $P_3 = 20\,$yr, since in this case the approximation of a strictly increasing chirp signal in frequency breaks down (cf.\,Eq.~\ref{eq: limit P3}) and our formula for the Doppler shift becomes invalid.   
}
\label{fig:relative error}
\end{figure}

In order to quantify the uncertainties with which we can constrain the parameters of the third body, we simulate a triple system, the parameters of which are summarized in Table.\,\ref{tab:parameters triple}. 
The particular values of the inner binary parameters were chosen very similar to the ones of \cite{Toubiana:2020cqv}.
This GW150914-like event would merge in 8 years observation time.  
The SNR associated to the event is 13.75. 

An example of an analysis of a triple system with an outer period of 33 years can be found in Figure \ref{fig:example result 10^8}. In accordance with expectations, for such low period the mass $m_3$ is measured with high precision: here to $\sim20\%$. 
However, we also find that the probability distribution is not centered around the true values of the triple system parameters, even though we neglected the noise contribution to the signal. This is because we estimate the \textit{posterior distribution} which includes the prior on the triple parameters, and we have assumed a log uniform prior on $m_3$ for a quicker convergence of the chains.
This, in turn, moves the posterior away from the true mass $m_3$ and shifts it towards lower values. However, we checked that the given uncertainties do not depend on the choice of the prior\footnote{Unfortunately, for large periods the analysis did not converge if a uniform prior on $m_3$ was assumed which is the reason for which we chose a uniform-in-logarithm prior. }.

Figure~\ref{fig:example result 10^8} underpins the need of a full parameter estimation -- a Fisher matrix approach cannot capture the full posterior distribution.  
The shape of the posterior of $m_3$ and $\iota_3$ can be readily understood: the data constrains the observable $m_3 \sin^3(\iota_3)$. Inverting this gives $\iota_3 = \arcsin(c/m_3^{1/3})$, where $c = \hat m_3 ^{1/3}\,\sin(\hat\iota_3)$ is a constant that will depend on the values of $\hat m_3$ and $\hat \iota_3$ that maximize the likelihood distribution. Indeed, this approximation describes well the posterior distribution between these two variables: the posterior samples are distributed along this thin line in parameter space. 
This explains the long convergence time for the naive choice of sampling parameters $m_3$ and $\iota_3$ -- many proposed samples are rejected, since they do not follow the precise trajectory in parameter space. 

Another interesting property visible from Figure~\ref{fig:example result 10^8} is the crucial role of the transverse Doppler term to estimate the mass of the SMBH $m_3$. Indeed, taking only into account the longitudinal Doppler term in the phase shift~\eqref{eq:FourierSpaceDoppler} would result in an \textit{exact} degeneracy when measuring $m_3$ and $\iota_3$, in the form $m_3 \sin^3 \iota_3 = \mathrm{Const}$. Thus, with longitudinal Doppler one would only be able to estimate a minimal mass for $m_3$, as the previous equation allows for solutions with arbitrarily large masses if $\iota_3$ is sufficiently close to zero. Estimating the mass of the SMBH with the longitudinal Doppler shift only would therefore result in a systematic discrepancy between this minimal mass $m_3^\mathrm{min} = \mathrm{Const}$ and the true mass $m_3$. We have also shown the value of $m_3^\mathrm{min}$ in Figure~\ref{fig:example result 10^8} in order to illustrate the importance of this systematic error on this particular example.

Note that for a period as short as $P_3=33$~yr, other three-body effects such as the Shapiro time delay discussed in~\cite{2022arXiv220508550S} should a priori be taken into account when modelling accurately the waveform of a three-body system. We do not include them here because they will likely bring very little improvement on the accuracy of parameter estimation. Indeed, their magnitude is always smaller than our degeneracy-breaking transverse Doppler term, as can be seen from Figure~\ref{fig:ParamSpace} and the discussion at the end of Section~\ref{sec:osculating}. However, it would be very interesting to build a complete waveform template taking into account all of these effects in the future.
As the period increases the Doppler effect caused by the central BH diminishes and we expect the uncertainty on the determination of the parameters of the outer orbit to increase, as well as correlations among different parameters as already discussed in Section~\ref{sec:gammas}. Eventually we reach a point where the convergence of the MCMC is very slow and the uncertainty on the SMBH mass $m_3$ becomes as large as the mass itself.

To study to which precision the mass of the SMBH $m_3$ can be measured we calculate the uncertainty of $m_3$ as it varies with the orbital period of the outer binary. Three cases are considered: a varying SMBH mass with values of $10^7\,\msun$, $10^8\,\msun$ and $10^9\,\msun$. 
We restrict the ranges of orbital periods to below $154$~yr, since the analysis did not converge for larger periods due to correlations among parameters.
In the following, we will denote the uncertainty corresponding to one standard deviation of $x$ as $\Delta x$. When the relative error of a variable $x$ is given, we calculate it as $\Delta x/ \text{median}(x)$.   

The uncertainty with which the SMBH mass can be constrained is dependent on $P_3$. For lower $P_3$ the inner binary is deeper in the potential well of the SMBH, and thus the frequency modulation of Eq.\,\eqref{eq: redshift third body} is stronger, resulting in a more precise measurement. From the aforementioned equation one can also see that the frequency shift is stronger for higher mass. This trend is also clearly visible in Figure \ref{fig:relative error}. 
The best (relative) constraint of $m_3$ is for the largest $m_3$ and the smallest outer period of 32 years considered here, with $\Delta m_3 / m_3\sim\,5\%$.
For a lower mass of $10^8\,\msun$ ($10^7\,\msun$), and a period of $P_3=20$\,yr, the relative uncertainty increases to $14\%$ ($34\%$). 
It is interesting to note that no other mechanism was proposed to accurately measure the mass of the SMBH from the GW of the inner binary for such high values of $P_3$: the Shapiro time delay and de Sitter precession effects already mentioned in the introduction were restricted to smaller period values (typically 1-2 years). Other astrophysical methods used to determine SMBH masses with similar precision are restricted to a sub-population of BH~\cite{2014SSRv..183..253P}, while our analysis covers all SMBH masses and could be used up to SMBH distances of a few GPc.

For larger periods, we generically find that the mass of the SMBH cannot be reliably estimated for a period $P_3 \gtrsim 150$ yrs. Note that the naive estimate in Figure~\ref{fig:ParamSpace} predicts a measurable transverse Doppler phase shift even for periods up to $P_3 \sim 1000$ yr. This difference highlights the important effect of correlations between parameters for a reliable parameter estimation based on this additional phase shift.


\section{Conclusions} \label{sec:conclusions}

With the future space-based GW detector LISA, we will detect stellar origin binaries many years before their coalescence. Depending on their astrophysical abundance we will observe 1 to 10 during the mission lifetime of LISA \cite{2022arXiv220508550S}. Some of these binaries are expected to be in orbit around a SMBH and the presence of this massive object leaves an observable imprint: the shift due to the motion of the inner binary in the direction of the line of sight to the observer (here termed the longitudinal effect). Besides this lowest-order (in post-Newtonian power-counting) Doppler shift, several other relativistic three-body effects can affect the waveform, for example
the Shapiro time delay due to the curved spacetime around the SMBH through which the signal has to propagate. 
These two effects allow one to measure the mass of the SMBH as was demonstrated in \cite{2022arXiv220508550S}. 

However, there is another effect whose characteristics have not yet been explored in the literature -- the signal also undergoes an additional frequency shift when it has a non-zero eccentricity. We refer to this as the \textit{transverse Doppler effect}.
In the work here-presented, we have explored how the mass of the SMBH can be measured when taking this latter effect into account. 
We found that the inclusion of the transverse Doppler shift, in the best case scenario considered here of a SMBH mass $m_3 = 10^9\,\msun$ and outer period $P_3  = 32$\,yr, allows for a measurement of the SMBH mass with a relative uncertainty of $5\%$. 
This uncertainty is dependent on the period of the inner binary around the central object $P_3$, as well as the mass of the central BH $m_3$. For a lower mass of $10^8\,\msun$ ($10^7\,\msun$), and a period of $P_3=20$\,yr, the relative uncertainty of $m_3$ increases to $14\%$ ($34\%$). For shorter periods, the uncertainty would be even lower, but we did not explore this range of parameter in our article since it would require to include many more three-body relativistic effects in order to obtain a proper modeling of the signal (the most important being that the frequency evolution can contain anti-chirping parts).
One advantage of our approach is that we can obtain a reasonable estimate of the SMBH mass $m_3$ even for outer periods much larger than the observation time of LISA. For example, we find that one can estimate $m_3$ with $30$\% uncertainty for a period $P_3=100$~yr and a mass $m_3=10^8 \, \msun$. To our knowledge, no other effect has been proposed in the literature to evaluate the SMBH mass from the GW signal for such large outer periods.

Our results are promising but also leave several directions open for future developments. For example, we chose here to focus on events which are observable by ground-based detectors so that parameters of the inner binary are independently measured, but it would be interesting to know how much our measurement degrades if we perform the analysis on an event not observed in ground-based interferometers. Another limitation of our study concerns the use of a restricted range of priors around the signal value: a true GW signal would require much more subtle data analysis techniques to be extracted from the noise, particularly since we are adding several parameters to be estimated on top of the usual two-body template. In this respect, choosing an efficient basis of non-degenerate parameters as we did in this article could prove crucial in order to analyze data in a reasonable time.

In the long-term, it would be desirable to obtain a waveform template taking into account all relativistic three-body effects necessary for modelling the signal of a binary BH in an AGN with enough precision for LISA data analysis. The identification of the relative importance of each of these effects is possible in a single consistent formalism with perturbative power-counting rules such as the one described in~\cite{Kuntz:2021ohi,Kuntz:2022onu}. Validating such a template against accurate numerical integration of the equations of motion would require a much more ambitious work than the one which we initiated in this article. 
Still, the leading order three-body effects in such a template would remain the longitudinal and transverse Doppler shifts investigated in this article.


\begin{acknowledgments}
We would like to thank Sylvain Marsat for helpful discussions.
This research has been partly supported by the Italian MIUR under contract 2017FMJFMW (PRIN2017). 
Numerical computations were performed on the DANTE platform, APC, France. KL was generously supported by the Fondation CFM pour la Recherche in Paris during his doctorate. 
We are grateful to the GDR-Ondes Gravitationnelles, supported by the CNRS, for providing a stimulating environment in which the authors initiated collaboration.
\end{acknowledgments}

\appendix


\section{Priors}
\label{app: priors}
\begin{figure}
    \centering
    \includegraphics[width = 0.8\textwidth]{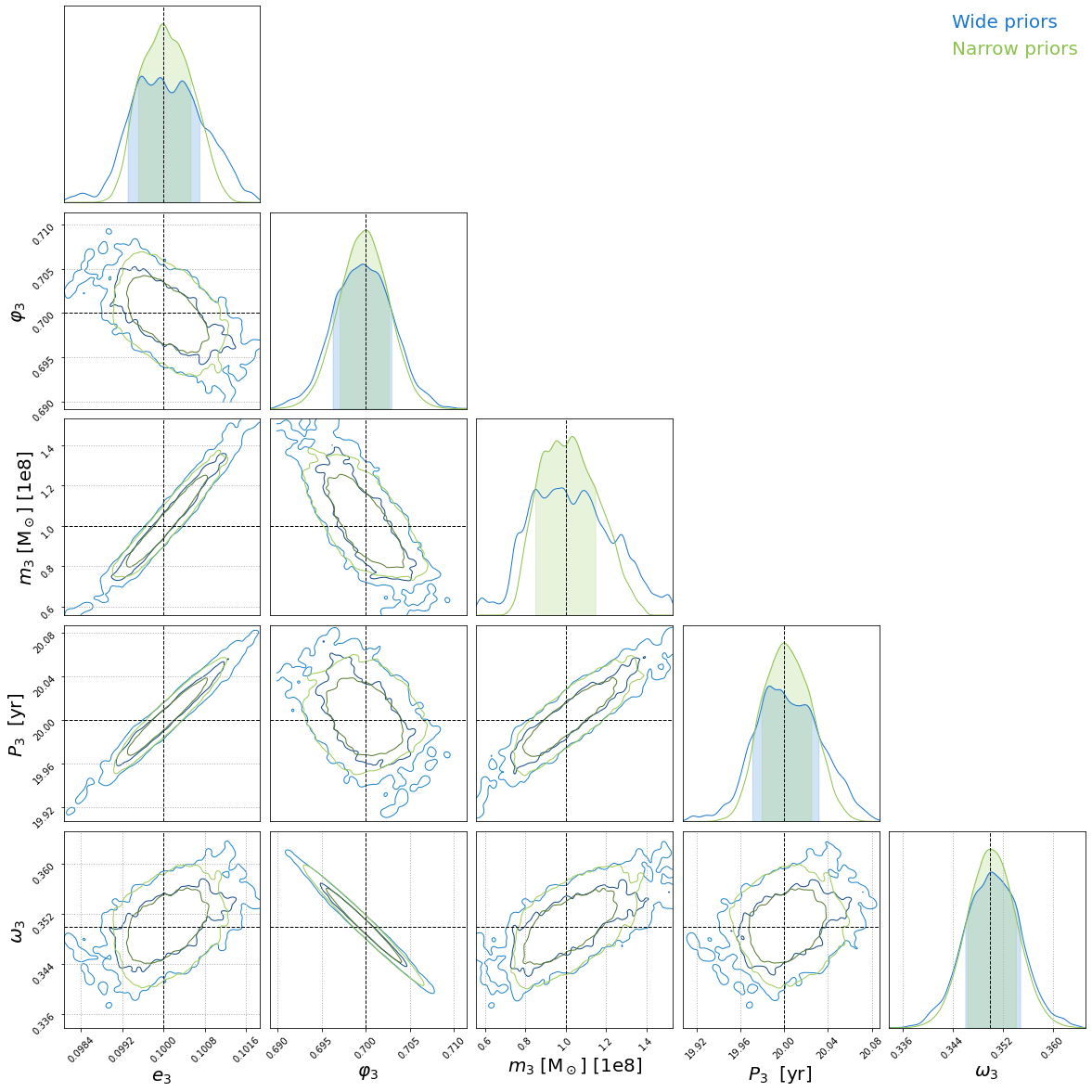}
    \caption{Result of the MCMC for a binary orbiting a $m_3 = 10^8\,\msun$ black hole, with a period of $P_3= 20$ years. The posterior under a wide (narrow) prior choice is marked with blue (green). The exact prior choices are summarized in Table~\ref{tab: wide and narrow priors}. Although the posterior under wide prior assumption is less well converged, we see that the two distributions are consistent. }
    \label{fig: comparison wide and narrow priors}
\end{figure}

We summarize in Table \ref{tab:priors parameters MCMC 10^7} the choices of prior for the events. For high periods of the outer period $P_3$, the prior range has to be carefully chosen: if the boundaries are too far apart, the chains do not converge within a reasonable computation time, if the prior domain is too small, the posterior has support outside of it and the uncertainties are underestimated. 
In order to assert that the choice of a limited prior range does not affect our results, we additionally perform a parameter estimation with larger prior range. We focus on a binary with an outer period of $P_3 = 20\,{\rm yr}$ around a SMBH with mass $m_3 = 10^8\,\msun$. Table \ref{tab: wide and narrow priors} compare the prior boundaries for a wide and narrow prior choice. 
Note that due to the increased parameter space in the wide prior case the number of MCMC steps was increased to 250.000. 
Fig.~\ref{fig: comparison wide and narrow priors} shows the resulting posterior distribution. The two posteriors are similar, and we conclude that the estimated uncertainties remain unaffected by the prior choice (assuming it does not cut-off the support of the true posterior).

\begin{table}[h!]
\renewcommand{\arraystretch}{1.5} 
\begin{center}
\begin{tabular}{|c| c c c c c c|} 
 \hline
 \multicolumn{7}{|c|}{ \textbf{Priors} }
  \\
 \hline
 \multicolumn{7}{c}{}\\[-1.0ex]
 \hline
  Injected value &  \multicolumn{6}{c|}{ $m_3 = 10^7 \msun$ }\\[0.5ex]
  \hline
  $P_3$ [yr] & $m_3$ [$10^7\,\msun$] & $m_3\,\sin^3(\iota_3)$ [$10^7\,\msun$]& $N$ [$10^7\,\msun$] & $e_3$ & $P_3$\,[yr] & $\varphi_3 - \omega_3$ \\
  \hline
20 & [0.2, 2] & [0.35, 0.39] & [0.236, 0.247] & [0.05, 0.15] & [19.7, 20.3] & [0.3, 0.4] \\
32 & [0.001, 2] & [0.35, 0.39] & [0.21, 0.255] & [0.08, 0.12] & [31.5, 32.6] & [0.24, 0.5] \\
54 & [0.001, 3] & [0.345, 0.39] & [0.21, 0.27] & [0.08, 0.125] & [51, 57] & [0.05, 0.7] \\
92 & [0.0001, 5] & [0.34, 0.4] & [0.19, 0.275] & [0.06, 0.14] & [86, 98] & [0.1, 0.65] \\
154 & [0.0001, 6] & [0.32, 0.41] & [0.19, 0.29] & [0.03, 0.16] & [142, 165] & [-0.2, 1] \\
\hline  
\multicolumn{7}{c}{}\\[-1.0ex]
\hline
Injected value &  \multicolumn{6}{c|}{ $m_3 = 10^8 \msun$ }\\[0.5ex]
  \hline
  $P_3$ [yr] & $m_3$ [$10^8\,\msun$] & $m_3\,\sin^3(\iota_3)$ [$10^8\,\msun$]& $N$ [$10^8\,\msun$] & $e_3$ & $P_3$\,[yr] & $\varphi_3 - \omega_3$ \\
  \hline
20 & [0.7, 1.4] & [0.36, 0.38] & [0.237, 0.247] & [0.05, 0.15] & [19.8, 20.3] & [0.3, 0.4] \\
32 & [0.1, 1.5] & [0.355, 0.38] & [0.21, 0.26] & [0.09, 0.11] & [31.5, 32.6] & [0.25, 0.48] \\
54 & [0.1, 1.6] & [0.34, 0.39] & [0.215, 0.26] & [0.08, 0.12] & [52, 57] & [0.1, 0.6] \\
92 & [0.01, 1.7] & [0.34, 0.39] & [0.21, 0.265] & [0.07, 0.135] & [87.5, 97] & [0.16, 0.5] \\
154 & [0.01, 3.6] & [0.35, 0.395] & [0.22, 0.26] & [0.07, 0.14] & [148, 160] & [0.05, 0.75] \\
\hline
\multicolumn{7}{c}{}\\[-1.0ex]
\hline
Injected value &  \multicolumn{6}{c|}{ $m_3 = 10^9 \msun$ }\\[0.5ex]
\hline
  $P_3$ [yr] & $m_3$ [$10^9\,\msun$] & $m_3\,\sin^3(\iota_3)$ [$10^9\,\msun$]& $N$ [$10^9\,\msun$] & $e_3$ & $P_3$\,[yr] & $\varphi_3 - \omega_3$ \\
  \hline
32 & [0.7, 1.2] & [0.36, 0.375] & [0.228, 0.25] & [0.095, 0.105] & [31.75, 32.3] & [0.25, 0.45] \\
54 & [0.6, 1.2] & [0.35, 0.38] & [0.22, 0.255] & [0.08, 0.12] & [52.8, 56] & [0.1, 0.5] \\
92 & [0.2, 1.5] & [0.335, 0.385] & [0.215, 0.255] & [0.087, 0.12] & [89, 96] & [0.1, 0.55] \\
154 & [0.02, 1.5] & [0.335, 0.385] & [0.22, 0.26] & [0.075, 0.125] & [151, 159] & [0.1, 0.5] \\
\hline
\multicolumn{7}{c}{}\\[-1.4ex]
\hline
Prior type & log uniform & log uniform & log uniform & uniform & uniform & uniform \\
 \hline
\end{tabular}
\caption{Priors for the source parameters, for varying SMBH mass $m_3$ and period of the outer binary $P_3$. For the true parameters of the source see table~\ref{tab:parameters triple}. To avoid cluttering, we use the definition $N=m_3 \sin^3(\iota_3) \sin^3(\varphi_3 + \omega_3)$.}
\label{tab:priors parameters MCMC 10^7}
\end{center}
\end{table}

\begin{table}[h!]
\renewcommand{\arraystretch}{1.5} 
\begin{center}
\begin{tabular}{|c| c c c c c c|} 
 \hline
Injected value &  \multicolumn{6}{c|}{ $m_3 = 10^8 \msun$, $P_3 = 20$\,yr }\\[0.5ex]
  \hline
  Variable & $m_3$ [$10^8\,\msun$] & $m_3\,\sin^3(\iota_3)$ [$10^8\,\msun$]& $N$ [$10^8\,\msun$] & $e_3$ & $P_3$\,[yr] & $\varphi_3 - \omega_3$ \\
  \hline
Prior choice (narrow) & [0.7, 1.4] & [0.36, 0.38] & [0.237, 0.247] & [0.05, 0.15] & [19.8, 20.3] & [0.3, 0.4] \\
Prior choice (wide) & [0.5, 2] & [0.32, 0.53] & [0.2, 0.27] & [0.03, 0.17] & [18.5, 21.5] & [0.25, 0.45] \\
Prior relative increase & 2.1 & 10 & 7 & 1.4 & 6 & 2 \\
\hline
Prior type & log uniform & log uniform & log uniform & uniform & uniform & uniform \\
 \hline
\end{tabular}
\caption{Priors for the runs with wide and narrow priors to study the impact of a narrow prior choice. For conciseness, we use again $N=m_3 \sin^3(\iota_3) \sin^3(\varphi_3 + \omega_3)$.}
\label{tab: wide and narrow priors}
\end{center}
\end{table}

\section{Impact of fixing the chirp mass}
\label{app: impact fixing chirp mass}

\begin{figure}
    \centering
    \includegraphics[width = 0.8\textwidth]{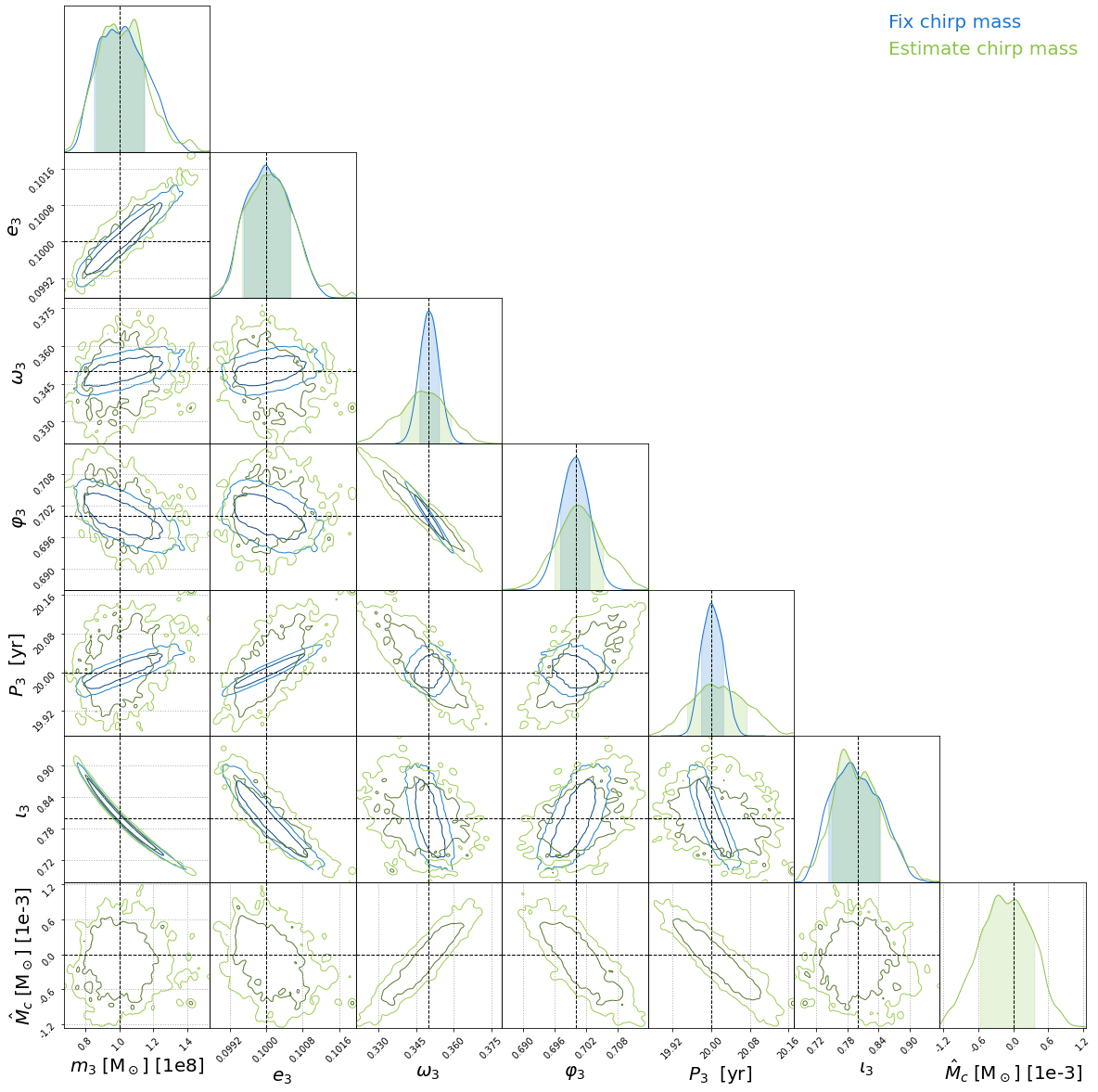}
    \caption{Result of the MCMC for a binary orbiting a $m_3 = 10^8\,\msun$ black hole, with a period of $P_3= 20$ years. The results when the chirp mass is fixed (blue) are compared to the results when the chirp mass is jointly estimated (green). The additional estimation of the chirp mass increases the uncertainties on the parameters $P_3$ and $\omega_3$ but leaves the marginal posterior of $m_3$ virtually unchanged. Note that we plot the chirp mass $\hat{M}_c = M_c - M_{c,{\rm true}}$, with $M_{c,{\rm true}}$ the true chirp mass.   }
    \label{comparison estimation of chirp mass}
\end{figure}

In the main body of this work, we have derived our results under the assumption that a ground-based detector network has inferred the parameters of the inner binary system. Of course, we can also ask the question of whether the estimated uncertainties change if we analyze data from LISA alone, i.e. estimating both the parameters of the inner and the outer orbits using LISA data. To gain insight into this more challenging situation, we now estimate the chirp mass of the inner binary in addition to the parameters of the outer orbit.
We study a binary that orbits a SMBH of $m_3 = 10^8\,\msun$ with a period of $20$ years. The other parameters are set to the values of Table.~\ref{tab:parameters triple}. 
Fig.~\ref{comparison estimation of chirp mass} summarizes the parameter estimation result -- although the uncertainties of some of the parameters (most notably $\omega_3$ and $P_3$) increase, the error of $m_3$ is unchanged.  
The parameters with increased uncertainties also exhibit the strongest correlations with the chirp mass.

\section{Convergence of MCMC}
\label{app: convergence}

To verify whether the chains of the MCMC analysis have converged, we compute the Gelman-Rubin criterion \cite{Gelman:1992zz}, denoted as $R$. If $R < 1.1$, the chains can be assumed to have converged \cite{Gelman:1992zz}. 
We use \texttt{arviz} \cite{arviz_2019} for the computation of $R$. 
From the results in Table \ref{tab: GR convergence criterion}, we see that the Gelman-Rubin criterion is satisfied for all parameters, indicating the convergence of all chains.


\begin{table}[h!]
\renewcommand{\arraystretch}{1.5} 
\begin{center}
\begin{tabular}{|c| c c c c c c c|} 
 \hline
 \multicolumn{8}{|c|}{ \textbf{Gelman-Rubin criterion} }
  \\
 \hline
 \multicolumn{8}{c}{}\\[-1.0ex]
 \hline
  Injected value &  \multicolumn{7}{c|}{ $m_3 = 10^7 \msun$ }\\[0.5ex]
  \hline
  $P_3$ & $m_3$ & $m_3\,\sin^3(\iota_3)$ & $N$  & $e_3$ & $P_3$ & $\varphi_3 - \omega_3$ & $\varphi $\\
  \hline
20 & 
1.026 & 1.021& 1.018& 1.006& 1.002& 1.001& 1.001 \\ 
32 & 
1.023 & 1.011& 1.012& 1.002& 1.004& 1.013& 1.002 \\ 
54 & 
1.027 & 1.024& 1.011& 1.013& 1.014& 1.006& 1.002 \\ 
92 & 
1.020 & 1.016& 1.017& 1.018& 1.017& 1.004& 1.003 \\ 
154 & 
1.026 & 1.007& 1.004& 1.005& 1.005& 1.004& 1.016 \\ 
\hline  
\multicolumn{8}{c}{}\\[-1.0ex]
\hline
Injected value &  \multicolumn{7}{c|}{ $m_3 = 10^8 \msun$ }\\[0.5ex]
  \hline
  $P_3$ & $m_3$ & $m_3\,\sin^3(\iota_3)$ & $N$  & $e_3$ & $P_3$ & $\varphi_3 - \omega_3$ & $\varphi $\\
  \hline
20 & 
1.029 & 1.024& 1.024& 1.011& 1.006& 1.003& 1.003 \\ 
32 & 
1.016 & 1.022& 1.024& 1.012& 1.016& 1.020& 1.006 \\ 
54 & 
1.014 & 1.021& 1.015& 1.016& 1.016& 1.002& 1.002 \\ 
92 & 
1.028 & 1.013& 1.011& 1.011& 1.011& 1.009& 1.007 \\ 
154 & 
1.018 & 1.019& 1.018& 1.021& 1.022& 1.003& 1.009 \\ 
\hline
\multicolumn{8}{c}{}\\[-1.0ex]
\hline
Injected value &  \multicolumn{7}{c|}{ $m_3 = 10^9 \msun$ }\\[0.5ex]
\hline
  $P_3$ & $m_3$ & $m_3\,\sin^3(\iota_3)$ & $N$  & $e_3$ & $P_3$ & $\varphi_3 - \omega_3$ & $\varphi $\\
  \hline
32 & 
1.021 & 1.017& 1.014& 1.007& 1.009& 1.014& 1.002 \\ 
54 & 
1.045 & 1.060& 1.033& 1.042& 1.041& 1.028& 1.007 \\ 
92 & 
1.019 & 1.020& 1.019& 1.021& 1.021& 1.017& 1.007 \\ 
154 & 
1.021 & 1.013& 1.009& 1.011& 1.012& 1.011& 1.006 \\ 
\hline
\end{tabular}
\caption{Gelman-Rubin convergence criterion for the MCMC analysis. To avoid cluttering, we use the notation $N=m_3 \sin^3(\iota_3) \sin^3(\varphi_3 + \omega_3)$.}
\label{tab: GR convergence criterion}
\end{center}
\end{table}



\bibliography{bib.bib}
\end{document}